\documentclass[floats, eqnum, showpacs, nofootinbib, 
,11pt
,preprint 
,eqsecnum 
]{revtex4-1}

\usepackage{color,graphicx}
\usepackage{amsfonts}
\usepackage{amssymb}
\begin{document}

\title{
High Energy Collision of Particles in the Vicinity of Extremal Black Holes in Higher Dimensions
\\
--{\it 
Ba\~nados-Silk-West Process as Linear Instability of Extremal Black Holes}--}

\author{
Naoki Tsukamoto,${}^{1,2}$\footnote{Electronic address:tsukamoto@rikkyo.ac.jp}
Masashi Kimura${}^{3,4}$\footnote{Electronic address:mkimura@yukawa.kyoto-u.ac.jp} 
and 
Tomohiro Harada${}^{1}$\footnote{Electronic address:harada@rikkyo.ac.jp}
}

\affiliation{
${}^{1}$Department of Physics, Rikkyo University, Tokyo 171-8501, Japan \\
${}^{2}$Center for Field Theory and Particle Physics \& Department of Physics, Fudan University, 200433 Shanghai, China \\
${}^{3}$Yukawa Institute for Theoretical Physics, Kyoto University, Kyoto 606-8502, Japan \\
${}^{4}$Osaka City University Advanced Mathematical Institute, 3-3-138 Sugimoto, Sumiyoshi, Osaka 558-8585, Japan
}
\date{\today}

\begin{abstract}
We study high energy particle collisions around higher dimensional black holes.
It is shown that the center of mass energy can be arbitrarily large 
in the vicinity of the event horizon like the
Ba\~nados, Silk and West~(BSW) process in four dimensions
if the following two conditions are satisfied: (i) the horizon is extremal and 
(ii) the parameters of 
either of the two colliding particles are fine-tuned, 
which is called a critical particle condition.
We also show that a test particle which satisfies the critical particle condition can reach the event horizon
from a distant region for a simple case.
Finally, we discuss the relation between the BSW process and 
the linear instability of test fields around 
extremal black holes, the latter of which
has been recently found by Aretakis~\cite{Aretakis:2011ha,Aretakis_2011,Aretakis:2011gz,Aretakis:2012ei,Aretakis:2012bm,Aretakis:2013dpa}.

\end{abstract}


\preprint{
RUP-13-12,
YITP-13-108,
OCU-PHYS-393, 
AP-GR-107
}

\maketitle


\section{Introduction}
\label{intro}

Ba\~nados, Silk and West pointed out that a
rapidly rotating black hole may act as a particle accelerator to arbitrarily high energy~\cite{Banados:2009pr}. 
We call this acceleration mechanism the Ba\~nados-Silk-West~(BSW) mechanism or BSW process.
There are several discussions on the BSW process~\cite{Banados:2009pr,Piran_Shaham_Katz_1975,Berti:2009bk,Jacobson:2009zg,Grib:2010dz,Lake:2010bq,
Wei:2010vca,Zaslavskii:2010aw,Harada:2010yv,Banados:2010kn,Kimura:2010qy,Zaslavskii:2010pw,
Patil:2011aw,Harada:2011xz,Patil:2011yb,Patil:2011uf,Harada:2011pg,Igata:2012js,Harada:2012ap,
McWilliams:2012nx,Nemoto:2012cq,Galajinsky:2013as,Zaslavskii:2013et,Nakao:2013uj}.
From a purely relativistic point of view, it is suggested in~\cite{Kimura:2010qy} that  
an arbitrarily high energy collision induced by the BSW process implies that 
an extremal Kerr black hole is linearly unstable.
If we consider the free-fall test particles with an arbitrarily small mass from the far region, 
we expect that the gravity induced by those test particles
will be well described by a linear perturbation around the background spacetime.
If one of the test particles has the fine-tuned angular momentum for given energy such that the BSW process works well,
an arbitrarily high energy particle collision can occur near the horizon.
In principle, the center of mass~(CM) energy can be as
large as the mass energy of the background black hole,
where such intense gravity cannot be described by a linear
perturbation around the background black hole.
Thus, it seems that we can construct a perturbation 
which can be described by a linear perturbation around an extremal Kerr black hole initially, 
but it cannot be after time evolution, so
that an extremal Kerr black hole is linearly unstable.

Recently, it is shown that the second-order derivative of a test field 
on the horizon of an extremal black hole blows up in late times
if the test field has a nonzero value on the horizon as an initial condition~\cite{Aretakis:2011ha,Aretakis_2011,Aretakis:2011gz,Aretakis:2012ei,Aretakis:2012bm,Aretakis:2013dpa}.
Unlike in the case of a nonextremal black hole, 
the test field can stay on the horizon for a long time and grow up.
In this sense, an extremal black hole has an instability on the horizon.

We would like to point out that there are some analogies between these two instabilities, 
both of which imply that the extremal black hole horizon is linearly unstable. 
First, it seems that both phenomena come from the vanishing surface gravity of the extremal horizon. 
In this case, an observer on the horizon does not feel gravitational force, and hence a particle or a field can stay in the vicinity of the horizon. 
Note that it is essential for the BSW process that there exists a geodesic particle which stays in the vicinity of the horizon. 
Next, a wave with sufficiently short wavelength behaves like a test particle. 
So, there should be a natural analogy between the dynamics of test particles and test fields. 
Therefore, we suggest that there is a close relationship between the BSW process and the instability of fields on extremal black holes.

According to the previous works~\cite{Banados:2009pr,Piran_Shaham_Katz_1975,Zaslavskii:2010aw,Wei:2010vca}, 
for an arbitrarily high energy
particle collision to occur by  the BSW process, we need to impose at least two conditions:
(i) the black hole horizon is extremal
and (ii) the parameters of either of the two colliding particles are fine-tuned. 
We can derive these two conditions by imposing that a particle can asymptote to the event horizon
in infinite proper time.
The condition (ii) is known as the critical particle condition which is a relation 
among energy, angular momentum and electric charge of the particle~\cite{Banados:2009pr,Piran_Shaham_Katz_1975,Zaslavskii:2010aw,Wei:2010vca}.
Since both the radial velocity and the acceleration of 
such a particle asymptote to zero 
as it approaches the horizon, i.e., 
\begin{eqnarray}
\frac{dr}{d\tau} \to 0~~(r \to r_+)
\label{eq:forcebalance1}
\\
\frac{d^2r}{d\tau^2} \to 0~~(r \to r_+),
\label{eq:forcebalance2}
\end{eqnarray}
where $\tau$ is the proper time of the particle and $r_+$ is the radius of the horizon,  
we might regard this condition as ``force balance'' conditions in an approach to the horizon.

It is proposed that the critical particle condition can be naturally realized 
if we consider an innermost stable circular orbit~(ISCO) particle.
In the case of a near extremal Kerr black hole, the radius of the ISCO 
is close to the horizon 
and such a particle satisfies $dr/d\tau \simeq 0$ and $d^2r/d\tau^2 \simeq 0$ in the vicinity of the horizon
which is similar to the conditions~(\ref{eq:forcebalance1})~and~(\ref{eq:forcebalance2}).
If we consider a collision between an ISCO particle 
and another free-fall particle around an extremal Kerr black hole,
an arbitrarily high energy particle collision can also
occur in the extremal limit as shown in~\cite{Harada:2010yv}.
We can expect that there is a relation between the approach of the ISCO to the horizon and the BSW process.

In this paper, to understand the nature of the BSW process deeper,
we study high energy particle collision around higher dimensional extremal black holes.
The effective potential for the radial coordinate $r$ roughly has the form of $-M/r^{D-3} + \Phi^{2}_{i}/r^2$, 
where $M$ is the mass of the black hole and $\Phi_{i}$ is the conserved angular momentum for a particle in a $D$-dimensional spacetime. 
The first term is the gravitational potential and the second term is the potential of centrifugal force. In higher dimensions $D \ge 5$, the power of gravitational force $-(D-3)$ is not greater than
that of centrifugal force $-2$. 
In that case, the effective potential does not have a local minimum; namely,
there is no stable circular orbit in contrast with four dimensions.
If the existence of ISCO is important for the BSW process, we can expect that
high energy particle collision cannot occur around higher dimensional extremal black holes.
On the other hand, recently, it was shown that there are linear instabilities of test fields on the horizon of higher dimensional extremal black holes~\cite{Murata:2012ct}.
If there exists an analogy between instabilities of the test field and the BSW process,
we can expect that a high energy particle collision can also occur in higher dimensions.

There is also a discussion for the case of charged particles 
around an extremal Reissner-Nordstr\"{o}m black hole, which can be considered as a similar system for
a free-fall test particle around an extremal Kerr black hole~\cite{Kimura:2010qy}.
In this case, we can understand the critical particle condition 
for a radially moving charged particle as the force balance between gravitational force and Coulomb force.
In the case of higher dimensions, we can expect high energy collision 
between two charged particles can also occur around an extremal charged black hole.
This is because gravitational force and Coulomb force have the same power law of radial distance in  
higher dimensional spacetimes, and the situation is the same as the four-dimensional case.

This paper is organized as follows.
In Sec.~\ref{secII}, 
we review the motion of a test particle around Myers-Perry black hole spacetimes 
and investigate a CM energy for the colliding two particles.
We show that the collision with an arbitrarily high CM energy can occur 
in the case of higher dimensions.
In Secs.~\ref{secIII} and \ref{secIV}, we discuss whether critical particles can reach the event horizon
by investigating effective potentials of test particles.
As a typical case, we also investigate the cases of five and six dimensions in detail.
In Sec.~\ref{secV}, we study the case of $D$-dimensional charged black holes.
Section~\ref{secVI} is devoted to the summary and discussion.
In this paper we use the units in which the light speed $c=1$.

\section{Myers-Perry black holes and particle motions}
\label{secII}
In this section, first, we review the motion of a test particle in the $D$-dimensional Myers-Perry black hole spacetime~\cite{Myers_Perry_1986,Vasudevan:2004mr,Emparan:2008eg}.
Next, we investigate the collision of two particles falling into the black hole and derive a CM energy of them.

We describe the Myers-Perry spacetime in $D~(\geq 4)$ dimensions in the Boyer-Lindquist coordinates 
with the spin parameters $\{a_{i}\}_{i=1,\ldots,n}$. 
Here $n=\lfloor D/2\rfloor$~$(\geq 2)$ 
where $\lfloor X \rfloor$ is the floor function or 
the greatest integer that is less than or equal to $X$.
For even dimensions, one of the spin parameter $a_{i}$ should vanish because of an extra unpaired spatial coordinate. 
We will adopt the convention that $a_{n}$ vanishes for even dimensions.
We can assume $a_{i}\geq 0$ for $i=1,\ldots,n$ without loss of generality. 

In this paper, for simplicity,
we focus on the case where the spin parameters $\{a_{i}\}$ take only one
value $a_{i}=a$ or take only two different 
values $a_{i}=a$ and $b$.
We adopt the convention 
$a_{i}=a$ for $i=1,\ldots,q$ and $a_{j+q}=b$ for $j=1,\ldots,p$,
where $q+p=n$.
We will assume that the parameter $a$ does not vanish.
Under the assumptions and convention, 
we can calculate the Hamilton-Jacobi equation and 
a CM energy of colliding two particles 
around a Myers-Perry black hole in a unified way
regardless of the parity of the spacetime dimensions.
In the even dimensional case, the black hole has $n-1$ equal angular momenta if we set $q=n-1$~(or $p=1$) because of the spin parameter $b= a_{n}=0$.
In the odd dimensional case, the black hole has $n$ equal angular momenta if we set $a=b$. 

The line element in the higher dimensional Myers-Perry black hole is explicitly expressed by the following form:
\begin{eqnarray}
ds^{2}
&=&-dt^{2}+\frac{Udr^{2}}{V-2M} +\rho^{2}d\theta^{2} \nonumber\\
&&+\frac{2M}{U} \left[ dt-a\sin^{2}\theta \sum^{q}_{i=1} \left( \prod^{q-i}_{k=1} \sin^{2}\alpha_{k} \right) \cos^{2}\alpha_{q-i+1}d\phi_{i} \right. \nonumber\\
&&\left. -b\cos^{2}\theta \sum^{p-\epsilon}_{j=1} \left( \prod^{p-j}_{k=1} \sin^{2}\beta_{k} \right) \cos^{2}\beta_{p-j+1}d\phi_{j+q} \right]^{2}\nonumber\\
&&+(r^{2}+a^{2})\sin^{2}\theta \sum^{q-1}_{i=1}\left( \prod^{i-1}_{k=1} \sin^{2}\alpha_{k} \right) d\alpha_{i}^{2}
+(r^{2}+b^{2})\cos^{2}\theta \sum^{p-1}_{j=1}\left( \prod^{j-1}_{k=1} \sin^{2}\beta_{k} \right) d\beta_{j}^{2}\nonumber\\
&&+(r^{2}+a^{2})\sin^{2}\theta \sum^{q}_{i=1} \left( \prod^{q-i}_{k=1} \sin^{2}\alpha_{k} \right) \cos^{2}\alpha_{q-i+1} d\phi_{i}^{2}\nonumber\\
&&+(r^{2}+b^{2})\cos^{2}\theta \sum^{p-\epsilon}_{j=1} \left( \prod^{p-j}_{k=1} \sin^{2}\beta_{k} \right) \cos^{2}\beta_{p-j+1} d\phi_{j+q}^{2},\qquad\quad 
\end{eqnarray}
where
$\epsilon$ is $1$ and $0$ for even and odd $D$ dimensions, respectively,
$\{\phi_{i}\}_{i=1,\ldots,n-\epsilon}$ is the azimuthal angular coordinates which take values in the range $0\leq \phi_{i} \leq 2\pi$,
$r$ is the radial coordinate,
$t$ is the temporal coordinate,
$\theta$ is the angular coordinate which takes a value in the range $0\leq \theta \leq \pi/2$,
$\{\alpha_{i}\}_{i=1,\ldots,q}$ and $\{\beta_{j}\}_{j=1,\ldots,p}$ are two sets of spherical polar coordinates, 
$M$ is the mass of the black hole and
\begin{eqnarray}
U&=&r^{\epsilon}\left( \frac{\sin^{2}\theta}{r^{2}+a^{2}}+\frac{\cos^{2}\theta}{r^{2}+b^{2}} \right) (r^{2}+a^{2})^{q}(r^{2}+b^{2})^{p-\epsilon}  \nonumber\\
&=&r^{\epsilon}\rho^{2} (r^{2}+a^{2})^{q-1}(r^{2}+b^{2})^{p-1-\epsilon}, \\
V&=&r^{\epsilon-2}(r^{2}+a^{2})^{q}(r^{2}+b^{2})^{p-\epsilon}, \\
\rho^{2}&=&r^{2}+a^{2}\cos^{2}\theta+b^{2}\sin^{2}\theta
\end{eqnarray}
and where
\begin{eqnarray}\label{eq:alpha_q_beta_p}
\alpha_{q}=\beta_{p}=0
\end{eqnarray}
and $\prod^{0}_{k=1}f_{k}=1$ for any function $f_{k}$.
If the equation $V - 2M =0$ has at least one positive root,
there exists an event horizon at $r = r_+$, where $r_+$ is the largest root.

For later convenience, we define the functions $\Delta(r)$, $\Pi(r)$ and $Z(r)$ as 
\begin{eqnarray}
&&\Delta(r) \equiv V-2M,\\
&&\Pi(r) 
\equiv \prod^{n-\epsilon}_{i=1} (r^{2}+a_{i}^{2})
=(r^{2}+a^{2})^{q}(r^{2}+b^{2})^{p-\epsilon}
\end{eqnarray}
and
\begin{eqnarray}
Z(r) \equiv (r^{2}+a^{2})(r^{2}+b^{2}),
\end{eqnarray}
respectively.
We note that the function $\Delta$ becomes zero at the event horizon $r = r_+$, i.e.
\begin{eqnarray}
\Delta(r_+) = 0.
\end{eqnarray}
These functions satisfy the following equations:
\begin{eqnarray}
&&U=\frac{r^{\epsilon}\rho^{2}\Pi}{Z},\\
&&\frac{V}{r^{\epsilon}\Pi}=\frac{1}{r^{2}},\\
&&\frac{V}{U}=\frac{(r^{2}+a^{2})(r^{2}+b^{2})}{r^{2}\rho^{2}}=\frac{Z}{r^{2}\rho^{2}}.
\end{eqnarray}
See Appendix~A for more details on the line element.

\subsection{Hamilton-Jacobi equation}
We discuss the motion of a test particle with the rest mass $m$ in the $D$-dimensional Myers-Perry black hole spacetime 
by using the Hamilton-Jacobi method.
We define the action $S=S(\lambda, x^{\mu})$ as a function of the coordinates $x^{\mu}$ and 
the parameter $\lambda = \tau/m$ for massive particle, 
where $\tau$ correspond to the proper time.
The Hamilton-Jacobi equation is given by
\begin{eqnarray}\label{eq:S6_Hamilton_Jacobi}
\frac{\partial S}{\partial \lambda}+H=0,
\end{eqnarray}
where 
\begin{eqnarray}
H\equiv \frac{1}{2} g_{\mu\nu}p^{\mu}p^{\nu}
\end{eqnarray}
is the Hamiltonian of a particle and $p_{\mu}$ is the conjugate momentum given by 
\begin{eqnarray}
p_{\mu}=\frac{\partial S}{\partial x^{\mu}}.
\end{eqnarray}
We can write $S$ in the following form of separation of variables because of the cyclic coordinates $t$ and $\phi_{i}$: 
\begin{eqnarray}
S 
=\frac{1}{2}m^{2}\lambda-Et+\sum^{q}_{i=1}\Phi_{i}\phi_{i}+\sum^{p}_{i=1}\Psi_{i}\phi_{q+i}
+S_{r}(r)+S_{\theta}(\theta)+\sum^{q-1}_{i=1}S_{\alpha_{i}}(\alpha_{i})  +\sum^{p-1}_{i=1}S_{\beta_{i}}(\beta_{i}), \qquad 
\end{eqnarray}
where $E\equiv-p_{t}$, $\Phi_{i}\equiv p_{\phi_{i}}$ for $i=1, \ldots, q$ and $\Psi_{j}\equiv p_{\phi_{q+j}}$ for $j=1, \ldots, p$ are constants 
and $S_{r}(r)$ and $S_{\theta}(\theta)$ are functions of the coordinates $r$ and $\theta$, respectively. 
For even dimensions, one of the conserved angular momenta $\Phi_{i}$ and $\Psi_{j}$ should vanish because of an extra unpaired spatial coordinate.
Here we have adopted the convention that $\Psi_{p}$ vanishes for even dimensions.
The constant
$E$ corresponds to the energy of the particle and $\{\Phi_{i}\}_{i=1,\ldots,q}$ and $\{\Psi_{j}\}_{j=1,\ldots,p}$ correspond to the 
angular momenta of the particle.

We can separate the Hamilton-Jacobi equation~(\ref{eq:S5_Hamilton_explitcit}), which is given in Appendix~B, 
for $r$ and $\theta$ as
\begin{eqnarray}\label{eq:S5_K_r}
K=(E^{2}-m^{2})r^{2}+\frac{2MZ}{r^{2}\Delta} \mathcal{E}^{2}(r)-\frac{\Delta Z}{r^{\epsilon}\Pi}\left( \frac{dS_{r}}{dr} \right)^{2}
+\frac{a^{2}-b^{2}}{r^{2}+a^{2}}J_{1}^{2}+\frac{-a^{2}+b^{2}}{r^{2}+b^{2}}L_{1}^{2}
\end{eqnarray}
and
\begin{eqnarray}\label{eq:S5_K_theta}
K=(m^{2}-E^{2})(a^{2}\cos^{2}\theta+b^{2}\sin^{2}\theta)
+\left( \frac{dS_{\theta}}{d\theta} \right)^{2}+\frac{J_{1}^{2}}{\sin^{2}\theta}+\frac{L_{1}^{2}}{\cos^{2}\theta},
\end{eqnarray}
respectively, 
where $K$, $J_{1}^{2}$ and $L_{1}^{2}$ are separation constants and we have defined the function $\mathcal{E}(r)$ as
\begin{eqnarray}
\mathcal{E}(r)\equiv E-\frac{a}{r^{2}+a^{2}}\sum^{q}_{i=1}\Phi_{i}-\frac{b}{r^{2}+b^{2}}\sum^{p}_{i=1}\Psi_{i}.
\label{mathcalE}
\end{eqnarray}
See Eqs.~(\ref{eq:S5_J1}) and~(\ref{eq:S5_L1}) for the definitions of $J_{1}^{2}$ and $L_{1}^{2}$, respectively, in Appendix~B.
From these equations, we obtain
\begin{eqnarray}
\frac{d\theta}{d\lambda} 
&=&
 \frac{dS_{\theta}}{d \theta}
=
\sigma_{\theta}\sqrt{\Theta},\\
\frac{dr}{d\lambda} 
&=&
\frac{dS_{r}}{dr}
=
\sigma_{r}\sqrt{R},
\label{eq:drdlambda}
\end{eqnarray}
with
\begin{eqnarray}\label{eq:S5_Theta}
\Theta (\theta)
\equiv (E^{2}-m^{2})(a^{2}\cos^{2}\theta+b^{2}\sin^{2}\theta)
+K-\frac{J_{1}^{2}}{\sin^{2}\theta}-\frac{L_{1}^{2}}{\cos^{2}\theta},
\end{eqnarray}
\begin{eqnarray}\label{eq:S5_R}
R(r)
\equiv
\frac{r^{\epsilon}\Pi}{\Delta Z} \left[ -K +(E^{2}-m^{2})r^{2}+\frac{2MZ}{r^{2}\Delta}\mathcal{E}^{2}(r) 
+\frac{a^{2}-b^{2}}{r^{2}+a^{2}}J_{1}^{2}
+\frac{-a^{2}+b^{2}}{r^{2}+b^{2}}L_{1}^{2}\right],
\end{eqnarray}
where
$\sigma_{\theta}=\pm 1$ and $\sigma_{r}=\pm 1$.
Note that we can choose the values of $\sigma_{\theta}$ and $\sigma_{r}$ independently
and $0\leq R$ and $0\leq \Theta$ must be satisfied for allowed particle motions.
From Eq.~(\ref{eq:S5_Theta}), the condition for $K$ is written as
\begin{eqnarray}\label{eq:S5_K_min}
(m^{2}-E^{2})(a^{2}\cos^{2}\theta+b^{2}\sin^{2}\theta) +\frac{J_{1}^{2}}{\sin^{2}\theta}+\frac{L_{1}^{2}}{\cos^{2}\theta} \leq K.\quad 
\end{eqnarray}

From the forward-in-time condition $dt/d\lambda \geq 0$,
we can see the condition 
\begin{eqnarray}
E+\frac{2MZ}{\Delta \rho^{2}r^{2}}\mathcal{E}(r) \geq 0
\end{eqnarray}
must be satisfied outside the event horizon.
By taking the limit $r \rightarrow r_{+}$, 
this condition yields
\begin{eqnarray}\label{eq:S5_forward_time}
\mathcal{E}(r_{+})=E-\frac{a}{r_{+}^{2}+a^{2}}\sum^{q}_{i=1}\Phi_{i}-\frac{b}{r_{+}^{2}+b^{2}}\sum^{p}_{i=1}\Psi_{i} \geq 0.
\end{eqnarray}

\subsection{Effective potential with respect to the radial coordinate}
Equation~(\ref{eq:drdlambda}) can be transformed to the familiar form as the energy equation, which is given by
\begin{eqnarray}\label{eq:S5_energy_equation}
\left( \frac{dr}{d\lambda} \right)^{2} +V_{\rm eff}(r,\theta)=0,
\end{eqnarray}
where $V_{\rm eff}(r,\theta)$ is the effective potential for a particle, which is defined as
\begin{eqnarray}\label{eq:S5_effective_potential}
V_{\rm eff}(r,\theta)
&\equiv&\frac{Z}{r^{2}\rho^{4}\Pi}T(r),
\end{eqnarray}
with 
\begin{eqnarray}\label{eq:S5_define_T}
T(r)
\equiv r^{2-\epsilon} \Delta
\left[ K +(m^{2}-E^{2})r^{2} 
-\frac{2MZ}{r^{2}\Delta}\mathcal{E}^{2}
-\frac{a^{2}-b^{2}}{r^{2}+a^{2}}J_{1}^{2} -\frac{b^{2}-a^{2}}{r^{2}+b^{2}}L_{1}^{2} \right]. \quad 
\end{eqnarray}
We obtain $V_{\rm eff} \rightarrow -E^{2}+m^{2}$ at large distances $r \rightarrow \infty$.
Therefore, a particle with $m^{2}-E^{2}>0$ is bound by the black hole or prohibited. 
We will discuss the effective potential in odd and even dimensions in detail in Secs. III and VI, respectively.

We define a critical particle as a particle which satisfies the condition 
\begin{eqnarray}\label{eq:S5_critical}
\mathcal{E}(r_{+})=E-\frac{a}{r_{+}^{2}+a^{2}}\sum^{q}_{i=1}\Phi_{i}-\frac{b}{r_{+}^{2}+b^{2}}\sum^{p}_{i=1}\Psi_{i}
=0.
\end{eqnarray}
In fact, this condition is obtained by solving the condition
\begin{eqnarray}
\left. \frac{dr}{d\lambda}\right|_{r=r_{+}}=0,
\end{eqnarray}
namely, the radial velocity becomes zero at the horizon.
For $b=0$ or $a=b$, 
the function $\mathcal{E}(r)$ in Eq.~(\ref{mathcalE}) for a critical particle can be written in a simple form as
\begin{eqnarray}
\mathcal{E}(r)=\frac{r^{2}-r_{+}^{2}}{r^{2}+a^{2}}E.
\end{eqnarray}

\subsection{Center of mass energy for colliding two particles}
We consider the collision of particles~$(1)$~and~$(2)$
whose rest masses and momenta
are given by $m_{(1)}$ and $p^{\mu}_{(1)}$ and $m_{(2)}$
and $p^{\mu}_{(2)}$, respectively.
The CM energy $E_{\rm CM}$ for the collision of the two particles at a collision point $r$ is obtained by 
\begin{eqnarray}
E_{\rm CM}^{2}
&=&-\left( p^{\mu}_{(1)}+p^{\mu}_{(2)} \right) \left( p_{(1) \mu}+p_{(2) \mu} \right) \nonumber\\
&=&m_{(1)}^{2}+m_{(2)}^{2}-2g^{\mu\nu}p_{(1)\mu}p_{(2)\nu}\nonumber\\
&=&m_{(1)}^{2}+m_{(2)}^{2}+2E_{(1)}E_{(2)} 
-\frac{2}{\rho^{2}}\sigma_{(1)\theta}\sigma_{(2)\theta}\sqrt{\Theta_{(1)}}\sqrt{\Theta_{(2)}}\nonumber\\
&&+\frac{4MZ}{r^{2}\rho^{2}\Delta}\mathcal{E}_{(1)}\mathcal{E}_{(2)}
-\frac{2\Delta Z}{r^{\epsilon} \rho^{2}\Pi}\sigma_{(1)r}\sigma_{(2)r}\sqrt{R_{(1)}}\sqrt{R_{(2)}}\nonumber\\
&&-\frac{2}{(r^{2}+a^{2})\sin^{2}\theta}\sum^{q}_{i=1}\frac{\Phi_{(1)i}\Phi_{(2)i}}{\prod^{q-i}_{k=1} \sin^{2}\alpha_{k}\cos^{2}\alpha_{q-i+1}}\nonumber\\
&&-\frac{2}{(r^{2}+b^{2})\cos^{2}\theta}\sum^{p}_{i=1}\frac{\Psi_{(1)i}\Psi_{(2)i}}{\prod^{p-i}_{k=1} \sin^{2}\beta_{k}\cos^{2}\beta_{p-i+1}}\nonumber\\
&&-\sum^{q-1}_{i=1}\frac{2\sigma_{(1)\alpha_{i}}\sigma_{(2)\alpha_{i}}\sqrt{A_{(1)i}}\sqrt{A_{(2)i}}}{(r^{2}+a^{2})\sin^{2}\theta \prod^{i-1}_{k=1} \sin^{2}\alpha_{k}}
-\sum^{p-1}_{i=1}\frac{2\sigma_{(1)\beta_{i}}\sigma_{(2)\beta_{i}}\sqrt{B_{(1)i}}\sqrt{B_{(2)i}}}{(r^{2}+b^{2})\cos^{2}\theta \prod^{i-1}_{k=1} \sin^{2}\beta_{k}},
\label{ecmgeneral}
\end{eqnarray}
where the subscripts $(1)$ and $(2)$ denote the quantities for the particles $(1)$ and $(2)$, respectively, 
and where $\sigma_{\alpha_{i}}=\pm 1$ and $\sigma_{\beta_{i}}=\pm 1$ are independent; 
see Eqs.~(\ref{eq:S5_An})-(\ref{eq:S5_Bp-1}) in Appendix~B for the definition of $A_{i}$ and $B_{i}$.
We can see that $E_{\rm CM}$ takes a finite value 
outside the event horizon for colliding particles with finite conserved quantities.
There is a possibility that
the center of mass energy $E_{\rm CM}$ can be arbitrarily 
large only in the limit where the collision point is 
arbitrarily close to the horizon.

On the event horizon, $\sigma_{(1)r}=\sigma_{(2)r}=-1$ must be satisfied.
Thus, we assume $\sigma_{(1)r}=\sigma_{(2)r}=-1$.
Using l'Hospital's rule with respect to $r^{2}$ in the near horizon limit $r\rightarrow r_{+}$, we obtain 
\begin{eqnarray}
\frac{4MZ}{r^{2}\rho^{2}\Delta}\mathcal{E}_{(1)}\mathcal{E}_{(2)}
-\frac{2\Delta Z}{r^{\epsilon} \rho^{2}\Pi}\sqrt{R_{(1)}}\sqrt{R_{(2)}}
=\frac{1}{\rho^{2}} \left( W_{(1)}\frac{\mathcal{E}_{(2)}}{\mathcal{E}_{(1)}}+W_{(2)}\frac{\mathcal{E}_{(1)}}{\mathcal{E}_{(2)}} \right),  
\end{eqnarray}
where
\begin{eqnarray}
W_{I}\equiv
K_{I} +(m_{I}^{2}-E_{I}^{2})r^{2}-\frac{a^{2}-b^{2}}{r^{2}+a^{2}}J_{I1}^{2}
+\frac{a^{2}-b^{2}}{r^{2}+b^{2}}L_{I1}^{2},\quad 
\end{eqnarray}
where $K_{I}$, $J_{I1}$ and $L_{I1}$ are $K$, $J_{1}$ and $L_{1}$ for
the particle $I=(1)$ or $(2)$. 
Here we have used Eq.~(\ref{eq:S5_forward_time}).
Thus, the CM energy for the collision of the two particles in the near horizon limit $r\rightarrow r_{+}$ is obtained by
\begin{eqnarray}\label{eq:CME}
\lim_{r \rightarrow r_{+}} E_{\rm CM}^{2}
&=&m_{(1)}^{2}+m_{(2)}^{2}+2E_{(1)}E_{(2)}
-\frac{2}{\rho^{2}}\sigma_{(1)\theta}\sigma_{(2)\theta}\sqrt{\Theta_{(1)}}\sqrt{\Theta_{(2)}}
+\frac{1}{\rho^{2}} \left( W_{(1)}\frac{\mathcal{E}_{(2)}}{\mathcal{E}_{(1)}}+W_{(2)}\frac{\mathcal{E}_{(1)}}{\mathcal{E}_{(2)}} \right) \nonumber\\
&&-\frac{2}{(r_{+}^{2}+a^{2})\sin^{2}\theta}\sum^{q}_{i=1}\frac{\Phi_{(1)i}\Phi_{(2)i}}{\prod^{q-i}_{k=1} \sin^{2}\alpha_{k}\cos^{2}\alpha_{q-i+1}}\nonumber\\
&&-\frac{2}{(r_{+}^{2}+b^{2})\cos^{2}\theta}\sum^{p}_{i=1}\frac{\Psi_{(1)i}\Psi_{(2)i}}{\prod^{p-i}_{k=1} \sin^{2}\beta_{k}\cos^{2}\beta_{p-i+1}}\nonumber\\
&&-\sum^{q-1}_{i=1}\frac{2\sigma_{(1)\alpha_{i}}\sigma_{(2)\alpha_{i}}\sqrt{A_{(1)i}}\sqrt{A_{(2)i}}}{(r_{+}^{2}+a^{2})\sin^{2}\theta \prod^{i-1}_{k=1} \sin^{2}\alpha_{k}}
-\sum^{p-1}_{i=1}\frac{2\sigma_{(1)\beta_{i}}\sigma_{(2)\beta_{i}}\sqrt{B_{(1)i}}\sqrt{B_{(2)i}}}{(r_{+}^{2}+b^{2})\cos^{2}\theta \prod^{i-1}_{k=1} \sin^{2}\beta_{k}}.
\end{eqnarray}
In the case of the black hole, this shows that a collision with an arbitrarily high CM energy can occur
if and only if one of the two particles is critical and the other is noncritical. 
For the prospect of collisions with high CM energy, it is important
to study whether critical particles can reach the event horizon.
Thus, we will discuss the effective potentials for critical particles in Secs.~III and IV.

\section{Myers-Perry black holes with equal spins in odd dimensions}
\label{secIII}

\subsection{Critical particles with general angular momenta}

In this section, we will investigate the effective potentials for 
the critical particles 
and test whether critical particles can reach the event horizon
in the odd dimensional Myers-Perry black hole spacetime with equal angular momenta $a=b$.

In this case, the effective potential for critical particles is given by
\begin{eqnarray}
V_{\rm eff}(r)
=\frac{T(r)}{r^{2}(r^{2}+a^{2})^{n}},
\end{eqnarray}
where
\begin{eqnarray}
T(r)
&=&
\left[ K +(m^{2}-E^{2})r^{2} \right] r^{2}\Delta(r) -2M(r^{2}-r_{+}^{2})^{2}E^{2}, \quad \quad
\label{todd}
\\
\Delta &=& \frac{1}{r^2} (r^2 + a^2)^n - 2M.
\end{eqnarray}
On the event horizon $r= r_{+}$, 
the effective potential vanishes, 
\begin{eqnarray}
V_{\rm eff}(r_+) &=& 0,
\label{eq:S5_horizon_odd_dimension}
\end{eqnarray}
because of the condition for the event horizon $\Delta(r_{+})=0$.
For the critical particle to reach the event horizon from $r > r_+$, 
the first derivative of the effective potential $V_{\rm
eff}^\prime(r_{+})$ should be zero or negative.
The function $V_{\rm eff}^\prime(r_{+})$ at the horizon becomes
\begin{eqnarray}
V_{\rm eff}^\prime(r_{+}) &=& 
\frac{T^\prime(r_{+})}{r_+^{2}(r_+^{2}+a^{2})^{n}},
\\
T^\prime(r_{+}) &=&
\left[ K +(m^{2}-E^{2})r_+^{2} \right] r_+^{2} \Delta^\prime (r_+).
\end{eqnarray}
Since $\Delta(r)$ takes a positive value outside the horizon, 
$\Delta^\prime (r_+)$ should be zero or positive.
In the case of $\Delta^\prime (r_+) > 0 $, 
$K +(m^{2}-E^{2})r_+^{2} \le 0 $ must be satisfied.
This case corresponds to the so-called multiple scattering scenario~\cite{Grib:2010dz}.
In this paper, we mainly focus on the case $\Delta^\prime (r_+) = 0$.
In this case, 
the function $V_{\rm eff}^\prime$ becomes zero at the
horizon, i.e., 
\begin{eqnarray}
V_{\rm eff}^\prime(r_{+}) &=& 0,
\label{eq:S5_extremal_odd_dimension}
\end{eqnarray}
and a particle asymptotes to the event horizon in infinite proper time.
From Eq.~(\ref{eq:S5_extremal_odd_dimension}), we obtain 
\begin{eqnarray}
&&a^{2}=r_{+}^{2}(n-1),
\label{aodddim}
\\
&&r_{+}=\left( \frac{2M}{n^{n}} \right)^{\frac{1}{2n-2}}.
\label{rpodddim}
\end{eqnarray}
We can see that the Myers-Perry black hole must have an extremal rotation.
Note that we can obtain the same conclusion for the case of even dimensions 
in a way similar to the above discussion.

From Eqs.~(\ref{eq:S5_horizon_odd_dimension}) and~(\ref{eq:S5_extremal_odd_dimension}),
we can see that both the radial velocity and the acceleration of the
test particle vanish at the horizon, i.e., 
\begin{eqnarray}\label{eq:dr_dlambda}
\left. \frac{dr}{d\lambda}\right|_{r=r_{+}}=-\sqrt{-V_{\rm eff}(r_{+})}=0,
\end{eqnarray}
and
\begin{eqnarray}\label{eq:ddr_dlambdadlambda}
\left. \frac{d^{2}r}{d\lambda^{2}}\right|_{r=r_{+}}=-\frac{1}{2}V_{\rm eff}^\prime(r_{+})=0.
\end{eqnarray}
This implies that the critical particle 
satisfies the force balance conditions~(\ref{eq:forcebalance1}) and~(\ref{eq:forcebalance2}) in the case of the extremal horizon.

It is useful to classify the effective potential with the sign of the second derivative $T^{\prime \prime}(r_{+})$ and $\alpha\equiv m^{2}/E^{2}$, where
\begin{eqnarray}
T^{\prime \prime}(r_{+})
=8M\frac{n-1}{n} \left[ K +\left( m^{2}-\frac{3n-1}{n-1}E^{2} \right) r_{+}^{2}  \right]. 
\end{eqnarray}
In the case $T^{\prime \prime}(r_{+})<0$ and $\alpha \leq 1$, 
$T(r)$, and hence $V_{\rm eff}(r)$ for a critical particle should be negative at least in the vicinity of the horizon; 
the critical particle is allowed to exist at large distances from the black hole and we refer to this class as class IA.
This class is most interesting because critical particles can reach from the infinity to the event horizon 
if $T(r)$ is negative everywhere outside the event horizon.
These critical particles can collide with a noncritical particle near the event horizon with an arbitrarily high CM energy.

In the case $T^{\prime \prime}(r_{+})<0$ and $\alpha >1$,
$T(r)$, and hence $V_{\rm eff}(r)$ for a critical particle should be negative at least in the vicinity of the horizon; 
the critical particle is always bounded by the black hole and we refer to this class as class IB.

In classes IA and IB , from the inequality~(\ref{eq:S5_K_min}) and $T^{\prime \prime}(r_{+})\leq 0$, $K$ should satisfy the condition
\begin{eqnarray}\label{eq:S5_K_inequality_all_dimension}
&&(m^{2}-E^{2})a^{2}+\frac{J_{1}^{2}}{\sin^{2}\theta} +\frac{L_{1}^{2}}{\cos^{2}\theta}
\leq  K \leq 
r_{+}^{2} \left( \frac{3n-1}{n-1}E^{2}-m^{2} \right). 
\label{krange1}
\end{eqnarray}

The case $T^{\prime \prime}(r_{+})=0$ is the marginal case, and we refer to this class as class II.
In four dimensions~($D=4$), this case corresponds to the last-stable-orbit collision~\cite{Harada:2010yv,Harada:2011xz}.

Suppose $T^{\prime \prime}(r_{+})>0$. 
Then, the critical particle cannot reach the horizon without multiple scattering because of the potential barrier and we refer to this class as class III.

For the critical particle to reach the horizon from the far region, 
the effective potential should satisfy $ V_{\rm eff}\le 0$ outside the horizon.
Though it is difficult to show it in general, we can prove it in the case of the
massless particle or a highly relativistic particle 
with $E^2 / m^2 \gg 1$.
Since the function $T$ is an increasing function of $K$ as seen in 
Eq.~(\ref{todd}), 
we only have to show $T \le 0$ for the maximum value of $K = K_{\rm max}$ in Eq.~(\ref{krange1}).
In this case, $T$ becomes 
\begin{eqnarray}
T|_{K=K_{\rm max}} &=& \left(r_+^2 \frac{3n-1}{n-1}E^2 - E^2 r^2\right) r^2 \Delta
-2 M (r^2 -r_+^2)^2 E^2.
\label{tkmax}
\end{eqnarray}
As shown in Appendix~\ref{vnegativeodd}
the function $T|_{K=K_{\rm max}} $ takes a negative value outside the horizon $r > r_+$.

\subsection{Critical particles with only one nonvanishing angular momentum}
For simplicity, we consider the case 
where the critical particle has only one nonvanishing 
angular momentum.
That is, we assume  
\begin{eqnarray}\label{eq:S5_One_conserved_angular_momentum}
\Psi_{i}=B_{j}=\Phi_{k}=A_{k}=0, \;   \Phi_{q} \neq 0, \; \cos^{2}\alpha_{1}=1
\end{eqnarray}
where $i=1$, $2$, \ldots, $p$, $j=1$, $2$, \ldots, $p-1$ and $k=1$, $2$, \ldots, $q-1$.
See Appendix~B for the definitions of $A_{i}$, $B_{i}$, $J_{i}^{2}$ and $L_{i}^{2}$.
From Eqs.~(\ref{eq:S5_Jq}), (\ref{eq:S5_Lp}), (\ref{eq:S5_An})-(\ref{eq:S5_Bp-1}) and~(\ref{eq:S5_One_conserved_angular_momentum}), 
we obtain 
\begin{eqnarray}\label{eq:J1_One_Conserved_Angular_Momentum}
J_{1}^{2}=\Phi_{q}^{2}
\end{eqnarray}
and 
\begin{eqnarray}\label{eq:J2_One_Conserved_Angular_Momentum}
J_{2}^{2}=J_{3}^{2}=\cdots=J_{q}^{2}=L_{1}^{2}=L_{2}^{2}=\cdots=L_{q}^{2}=0.
\end{eqnarray}
Thus, the critical particle must satisfy 
\begin{eqnarray}\label{eq:Critical_One_Conserved_Angular_Momentum}
J_{1}^{2}=\frac{(r_{+}^{2}+a^{2})^{2}}{a^{2}}E^{2}.
\end{eqnarray}
The above results~(\ref{eq:J1_One_Conserved_Angular_Momentum})-(\ref{eq:Critical_One_Conserved_Angular_Momentum}) are not only true in odd dimensions 
but also in even dimensions under the assumption~(\ref{eq:S5_One_conserved_angular_momentum}).

Under the assumption~(\ref{eq:S5_One_conserved_angular_momentum}), from the inequality~(\ref{eq:S5_K_inequality_all_dimension}), 
we can find $K$ which satisfies the inequality~(\ref{eq:S5_K_inequality_all_dimension}) if the following inequality is satisfied:  
\begin{eqnarray}
\left(\alpha-\frac{n+1}{n-1}\right) \sin^{2}\theta  \leq -\frac{n}{n-1}.
\end{eqnarray}
Therefore, we can find $K$ if $\alpha$ satisfies $\alpha\leq (n+1)/(n-1)$ and $\theta$ satisfies
\begin{eqnarray}\label{eq:S5_collision_belt_all_dimension}
\sin^{2}\theta \geq \frac{n}{n(1-\alpha)+1+\alpha}.
\end{eqnarray}
On the equatorial plane $\sin \theta=1$, we will find $K$ in all odd dimensions if $\alpha$ satisfies
$\alpha \leq 1/(n-1)$.
This shows that the particle must be massless or highly relativistic for $K$ to exist in the large dimension limit $n\rightarrow \infty$.

\subsection{Five-dimensional case}
As the simplest example of odd dimensional cases, we will consider the motion of the critical particle 
in the extremal five-dimensional Myers-Perry black hole spacetime~($a+b=\sqrt{2M}$) with equal angular momenta~($a=b$)~\cite{Myers_Perry_1986,Vasudevan:2004mr,Frolov:2002xf,Frolov:2003en,Aliev:2004ec,Emparan:2008eg}
and investigate the BSW effect there.
This case is given by setting $p=q=1$ and $n=2$. 
The explicit form of the line element of the Myers-Perry black hole spacetime is given in Appendix~C. 

The event horizon is obtained by
\begin{eqnarray}\label{eq:S5_5dim_horizon_extremal_degenerate}
r_{+}=a=\sqrt{M/2}.
\end{eqnarray}
From Eqs.~(\ref{eq:S5_critical}) and~(\ref{eq:S5_5dim_horizon_extremal_degenerate}) and $q=p=1$, we get
\begin{eqnarray}\label{eq:S5_5dim_critical_extremal_degenerate}
2aE=\Phi_{1}+\Psi_{1}.
\end{eqnarray}
From Eqs.~(\ref{eq:alpha_q_beta_p}), (\ref{eq:S5_5dim_critical_extremal_degenerate}), (\ref{eq:S5_Jq}) and~(\ref{eq:S5_Lp})  and $q=p=1$,   
we obtain
\begin{eqnarray}
J_{1}^{2}=\Phi_{1}^{2}
\end{eqnarray}
and 
\begin{eqnarray}
L_{1}^{2}=\Psi_{1}^{2}=(2aE-\Phi_{1})^{2}.
\end{eqnarray}

The effective potential of the critical particle is given by
\begin{eqnarray}\label{eq:S6_effective_potential}
V_{\rm eff}=\frac{T(r)}{(r^{2}+a^{2})^{2}r^{2}},
\end{eqnarray}
where
\begin{eqnarray}\label{eq:S6_effective_potential}
T(r)\equiv (r^{2}-a^{2})^{2} \left[ K+(m^{2}-E^{2})r^{2}-4a^{2}E^{2} \right].
\end{eqnarray}
Figure 1 shows the examples of the effective potential $V_{\rm eff}(r)$. 
In  Fig. 1, the curves of classes IA, IB, II and III 
correspond to the cases $T^{\prime \prime}(r_{+})<0$ and $\alpha \leq 1$, $T^{\prime \prime}(r_{+})<0$ and $\alpha >1$, $T^{\prime \prime}(r_{+})=0$ and $T^{\prime \prime}(r_{+})<0$, respectively. 

\begin{figure}[htbp]
\begin{center}
\includegraphics[width=87mm]{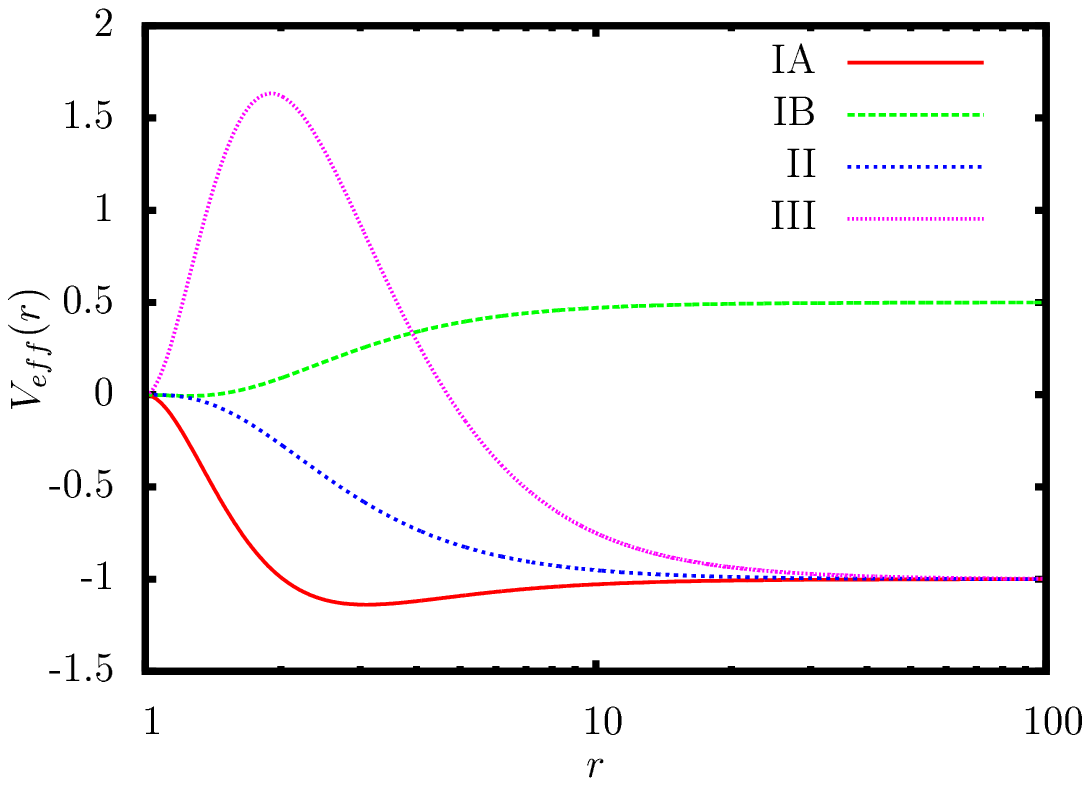}
\end{center}
\caption{The examples of the effective potential $V_{\rm eff}(r)$ for a
 critical particle  
in the five-dimensional and extremal Myers-Perry black hole spacetime with equal angular momenta.
The solid~(red), dashed~(green), dot-spaced~(blue) and dotted~(purple) curves denote
the effective potential for the critical particle in classes IA~($a=m=K=1$, $E=\sqrt{2}$), 
IB~($a=m=K=1$, $E=1/\sqrt{2}$), 
II~($a=m=1$, $K=9$, $E=\sqrt{2}$) and 
III~($a=m=1$, $K=30$, $E=\sqrt{2}$), respectively.
}
\label{fig:S6_two}
\end{figure}

We consider classes IA and IB and therefore assume $T^{\prime \prime}(r_{+}) \leq 0$, where
\begin{eqnarray}
T^{\prime \prime}(r_{+})= 8a^{2} \left[ K+(m^{2}-5E^{2})a^{2} \right].
\end{eqnarray}
For five dimensions, the inequality~(\ref{eq:S5_K_inequality_all_dimension}) is described as
\begin{eqnarray}\label{eq:S6_K_inequality}
(m^{2}-E^{2})a^{2}+\frac{\Phi^{2}_{1}}{\sin^{2}\theta}+\frac{(2aE-\Phi_{1})^{2}}{\cos^{2}\theta}
\leq K 
\leq (5E^{2}-m^{2})a^{2}.
\end{eqnarray}

In the case $\sin^{2}\theta=1$ with $2aE=\Phi_{1}$ or $\sin^{2}\theta=0$ with $\Phi_{1}=0$, 
the inequality~(\ref{eq:S6_K_inequality}) is satisfied and we can find $K$ which has a finite value if and only if $0\leq \alpha \leq 1$.
On the other hand, in the case $\sin^{2}\theta=1$ with $2aE\neq \Phi_{1}$ or $\sin^{2}\theta=0$ with $\Phi_{1}\neq 0$,
we cannot get $K$.

Next, we consider the case where $\sin^{2}\theta$ is neither $0$ nor $1$.
In this case, $K$ which satisfies the inequality~(\ref{eq:S6_K_inequality}) exists
if the following quadratic inequality for $\sin^{2}\theta$ is satisfied:
\begin{eqnarray}\label{eq:S6_sin_square_inequality}
2(3-\alpha)\sin^{4}\theta+2(\alpha-1-2\beta)\sin^{2}\theta+\beta^{2}\leq 0,
\end{eqnarray}
where $\beta \equiv \Phi_{1}/(aE)$. 
The inequality (\ref{eq:S6_sin_square_inequality}) is expressed by 
\begin{eqnarray}\label{eq:S6_sin_square_inequality2}
(\beta-2\sin^{2}\theta)^{2}+2\sin^{2}\theta (1-\sin^{2}\theta)(\alpha-1) \leq 0.
\end{eqnarray}
For $\alpha >1$, we easily see that the inequality~(\ref{eq:S6_sin_square_inequality2}) is not satisfied.
For $0 \leq \alpha \leq 1$,
the inequality (\ref{eq:S6_sin_square_inequality}) has the solution
$\theta$ satisfying $\zeta_{-} \leq \sin^{2}\theta \leq  \zeta_{+}$, where
\begin{eqnarray}
\zeta_{\pm} \equiv \frac{1-\alpha+2\beta \pm \sqrt{(1-\alpha)[1-\alpha+2\beta(2-\beta)]}}{2(3-\alpha)} \nonumber\\
\end{eqnarray}
and the range of $\beta$ is given by
\begin{eqnarray}
1-\frac{\sqrt{6-2\alpha}}{2} \leq \beta \leq 1+\frac{\sqrt{6-2\alpha}}{2}.
\end{eqnarray}
We plot $\zeta_{\pm}$ in Fig. 2. 

\begin{figure}[htbp]
\begin{center}
\includegraphics[width=87mm]{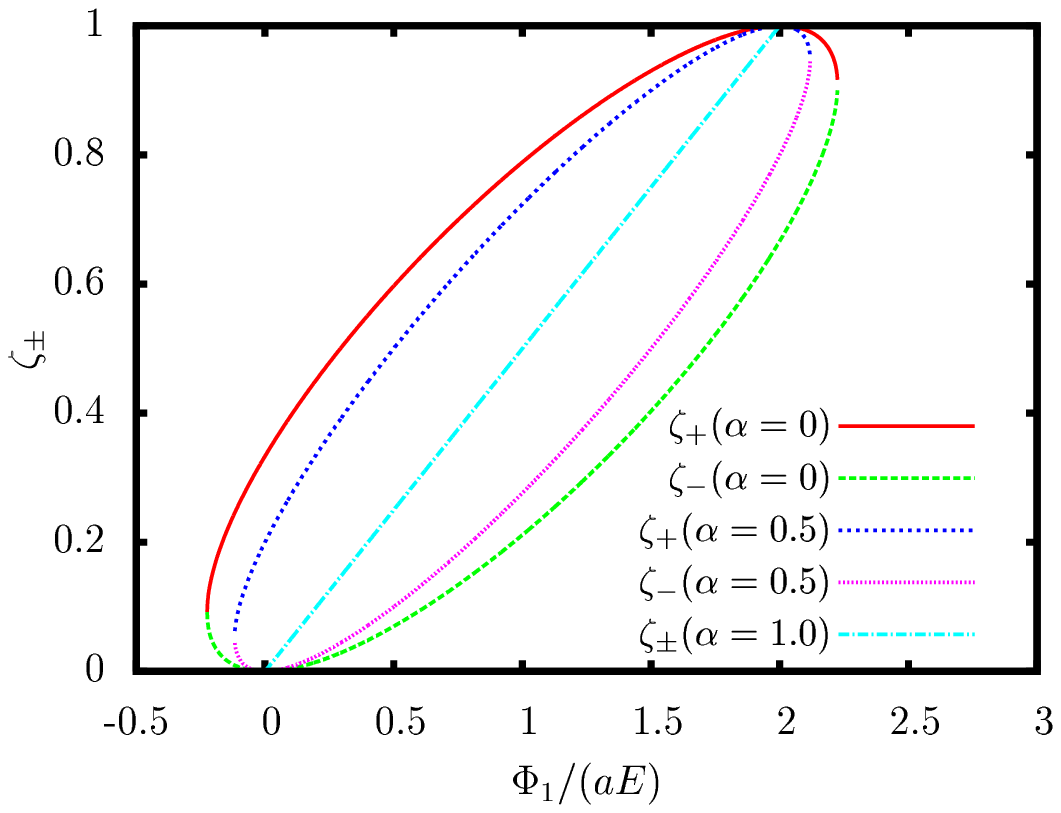}
\end{center}
\caption{$\zeta_{\pm}$ versus $\beta \equiv \Phi_{1}/(aE)$ for
 a critical particle in the five-dimensional and extremal Myers-Perry black hole spacetime with equal angular momenta.
The solid~(red), dashed~(green), dot-spaced~(blue), dotted~(purple) and dash-dotted~(cyan) curves denote
$\zeta_{+}$ and $\zeta_{-}$ for $\alpha=0$, $\zeta_{+}$ and $\zeta_{-}$ for $\alpha=0.5$ 
and $\zeta_{\pm}$ for $\alpha=1$, respectively.
In the region $\zeta_{-} \leq \sin^{2}\theta \leq \zeta_{+}$ including $\zeta_{-}=0$ and $\zeta_{+}=1$,
collisions with arbitrarily high CM energy occur.
}
\label{fig:S6_one}
\end{figure}

Therefore, we can find $K$ in the region $\zeta_{-} \leq \sin^{2}\theta \leq  \zeta_{+}$ 
for  $0 \leq \alpha \leq 1$ and $1-\sqrt{6-2\alpha}/2 \leq \beta \leq 1+\sqrt{6-2\alpha}/2$.
Collisions with an arbitrarily high CM energy occur within this region.

\section{Myers-Perry black holes with equal spins in even dimensions}
\label{secIV}

\subsection{Critical particles with general angular momenta}

In this section we discuss the effective potential for a critical particle in the even dimensional and extremal Myers-Perry black hole spacetime 
with the equal angular momenta $a_{i}=a$ for $i=1, 2,\ldots, n-1$ and $a_{n}=b=0$.
In this case we have to set $p=1$ and $q=n-1$.
Notice $L_{p}^{2}=L_{1}^{2}=0$ in the even dimensional case because of the convention $\Psi_{p}=\Psi_{1}=0$ and Eq.~(\ref{eq:S5_Lp}). 
The effective potential for a critical particle is given by
\begin{eqnarray}
V_{\rm eff}(r)
=\frac{T(r)}{\rho^{4}(r^{2}+a^{2})^{n-2}},
\end{eqnarray}
where
\begin{eqnarray}
T(r)
=
\left[ K +(m^{2}-E^{2})r^{2} -\frac{a^{2}}{r^{2}+a^{2}}J_{1}^{2} \right] r\Delta 
-2Mr\frac{(r^{2}-r_{+}^{2})^{2}}{r^{2}+a^{2}}E^{2}
\label{tevendim}
\end{eqnarray}

In the same way as for odd dimensions, we can show that the black hole must have the extremal rotation
so that critical particles reach the event horizon, where the force
balance conditions~(\ref{eq:forcebalance1}) and~(\ref{eq:forcebalance2}) hold.

From the condition for the event horizon $r_{+}^{2}\Delta(r_{+})=0$
and the condition for the extremal rotation $\left(r_{+}^{2}\Delta(r_{+})\right)^{\prime}=0$,
we obtain
\begin{eqnarray}
\label{eq:S5_a_even_p1}
&&a^{2}=r_{+}^{2}(2n-3)
\\
\label{eq:S5_horizon_even_p1}
&&r_{+}=(2M)^{\frac{1}{2n-3}}(2n-2)^{\frac{-n+1}{2n-3}}.
\end{eqnarray}

We concentrate on classes IA and IB, and hence $T^{\prime \prime}(r_{+})\leq 0$, where
\begin{eqnarray}
T^{\prime \prime}(r_{+})
=\left[ K +\left( m^{2}-\frac{2n+1}{2n-3}E^{2} \right) r_{+}^{2}-\frac{2n-3}{2n-2}J_{1}^{2}\right]
(2M)^{\frac{2n-4}{2n-3}}(2n-2)^{\frac{-n+2}{2n-3}}2(2n-3). \qquad 
\end{eqnarray}

From the inequality~(\ref{eq:S5_K_min}) and $T^{\prime \prime}(r_{+})\leq 0$, $K$ should satisfy the condition 
\begin{eqnarray}\label{eq:S5_inequarity_K}
(m^{2}-E^{2})a^{2}\cos^{2}\theta+\frac{J_{1}^{2}}{\sin^{2}\theta} 
\leq K \leq 
\left( \frac{2n+1}{2n-3}E^{2}-m^{2} \right) r_{+}^{2}+\frac{2n-3}{2n-2}J_{1}^{2}.
\label{krangeeven}
\end{eqnarray}

We can find $K$ if we get $\sin^{2}\theta$ which satisfies the inequality 
\begin{eqnarray}\label{eq:S5_inequality_even}
(1-\alpha)(2n-3)\sin^{4}\theta 
+\left[ 2\alpha(n-1)-(2n-3)-\frac{2n+1}{2n-3} -\frac{2n-3}{2n-2}\frac{J_{1}^{2}}{r_{+}^{2}E^{2}} \right] \sin^{2}\theta 
+\frac{J_{1}^{2}}{r_{+}^{2}E^{2}}\leq 0. \qquad 
\end{eqnarray}

\subsection{Critical particles with only one nonvanishing angular momentum}
We assume the one conserved angular momentum case~(\ref{eq:S5_One_conserved_angular_momentum}) for simplicity.
Under the assumption~(\ref{eq:S5_One_conserved_angular_momentum}), 
we also obtain Eqs.~(\ref{eq:J1_One_Conserved_Angular_Momentum})-(\ref{eq:Critical_One_Conserved_Angular_Momentum}) in even dimensions 
and the inequality~(\ref{eq:S5_inequality_even}) yields 
\begin{eqnarray}\label{eq:S5_inequality_even_one_conserved_angular_momentum}
(1-\alpha)(2n-3)^{2}\sin^{4}\theta 
+2\left[ \alpha(n-1)(2n-3)-4n^{2}+10n-8\right] \sin^{2}\theta 
+4(n-1)^{2} \leq 0. \qquad 
\end{eqnarray}

For $\alpha=0$, from the inequality~(\ref{eq:S5_inequality_even_one_conserved_angular_momentum}), we obtain the condition
\begin{eqnarray}
(2n-3)^{2}\sin^{4}\theta +2(-4n^{2}+10n-8)\sin^{2}\theta 
+4(n-1)^{2} \leq 0.
\end{eqnarray}
Thus, we can find $K$ for massless particles such as 
photons or highly relativistic massive particles with $E/m \gg 1$ 
in all even dimensions in the region where $\sin \theta$ satisfies the condition 
\begin{eqnarray}
\eta_{-} \leq \sin^{2}\theta \leq 1 \; (\leq \eta_{+}),
\end{eqnarray}
where
\begin{eqnarray}
\eta_{\pm}\equiv \frac{\left(\sqrt{4n^{2}-10n+7}\pm 1\right)^{2}}{(2n-3)^{2}}.
\end{eqnarray}

On the equatorial plane $\sin^{2}\theta=1$, from the inequality~(\ref{eq:S5_inequality_even_one_conserved_angular_momentum}),
we see that $K$ will be found in all even dimensions if $\alpha\leq 3/(2n-3)$ is satisfied.
Collisions with an arbitrarily high CM energy occur within this region.

In the case of massless or highly relativistic 
massive particles with $E/m\gg 1$, 
we can show $V_{\rm eff} < 0$ outside the horizon $r >r_+$, which is
similar to the case of odd dimensions.
The proof is given in Appendix~\ref{vnegativeeven}

\subsection{Six-dimensional case}
As the simplest example of the higher and even dimensional case, we will consider the effective potential for a critical particle 
in the six-dimensional and extremal Myers-Perry black hole spacetime with the equal angular momenta.
We have to set $q=2$, $p=1$ and $n=3$.
For simplicity, we assume $\Phi_{1}=A_{1}=0$ and $\Phi_{2}\neq 0$.
See Eq.~(\ref{eq:S5_An}) for the definition of $A_{1}$ in Appendix~B.
The inequality~(\ref{eq:S5_inequality_even}) is reduced to
\begin{eqnarray}\label{eq:S5_inequality_6d}
(-E^{2}+m^{2})\sin^{4}\theta +\left( \frac{16}{9}E^{2}+\frac{J_{1}^{2}}{M^{\frac{2}{3}}}-\frac{4}{3}m^{2} \right) \sin^{2}\theta 
-\frac{4}{3}\frac{J_{1}^{2}}{M^{\frac{2}{3}}}\geq 0.
\end{eqnarray}
Equations~(\ref{eq:S5_Jq}) and~(\ref{eq:alpha_q_beta_p}) imply 
\begin{eqnarray}\label{eq:S5_J_Phi_6d}
J_{1}^{2} =\frac{\Phi_{2}^{2}}{\cos^{2}\alpha_{1}} \geq \Phi_{2}^{2}.
\end{eqnarray}
Equations~(\ref{eq:S5_a_even_p1}) and~(\ref{eq:S5_horizon_even_p1}) imply
\begin{eqnarray}\label{eq:S5_a_horizon_M_6d}
a^{2}=3r_{+}^{2}=\frac{3}{4}M^{\frac{2}{3}}.
\end{eqnarray}
From Eqs.~(\ref{eq:S5_critical}) and~(\ref{eq:S5_a_horizon_M_6d}), the critical particle should satisfy 
\begin{eqnarray}\label{eq:S5_critical_6d}
\Phi_{2}=\frac{2}{\sqrt{3}}M^{\frac{1}{3}}E.
\end{eqnarray}

The effective potential for the critical particle is explicitly expressed by 
\begin{eqnarray}
\frac{\rho^{4}}{r^{4}}V_{\rm eff}(r,\theta)
&=&\frac{1}{r^{4}(r^{2}+a^{2})}\left\{ \left[ K+(m^{2}-E^{2})r^{2}-\frac{a^{2}J_{1}^{2}}{r^{2}+a^{2}} \right] \right. \nonumber\\
&&\left.  \left[ (r^{2}+a^{2})^{2}-\frac{16a^{3}r}{3\sqrt{3}} \right]
-\frac{16a^{3}r}{27\sqrt{3}}\frac{(3r^{2}-a^{2})^{2}E^{2}}{r^{2}+a^{2}}  \right\}. \qquad  
\end{eqnarray}
Figure 3 shows the examples of the effective potential $\rho^{4}V_{\rm eff}/r^{4}$.
\begin{figure}[htbp]
\begin{center}
\includegraphics[width=87mm]{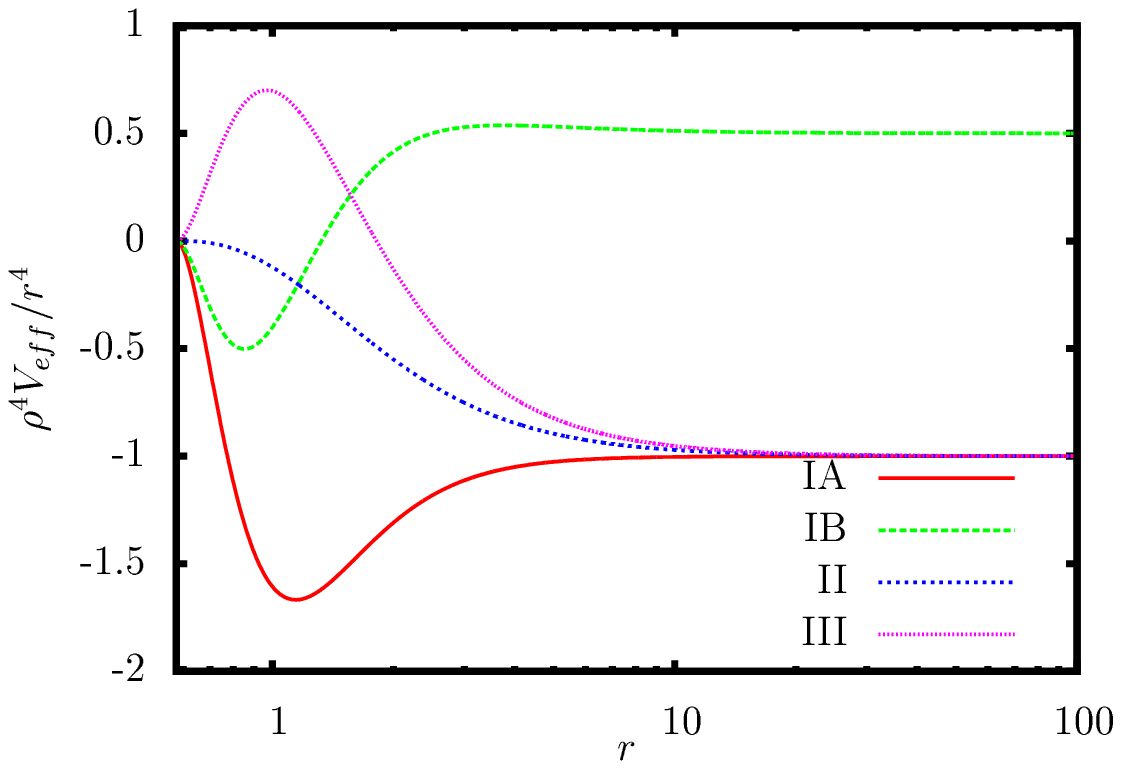}
\end{center}
\caption{The examples of the effective potentials $V_{\rm eff}(r,\theta)$ for critical particles  
in the six-dimensional and extremal Myers-Perry black hole spacetime with equal angular momenta.
The solid~(red), dashed~(green), dot-spaced~(blue) and dotted~(purple) curves denote
the effective potentials for the critical particles of classes IA~($a=m=K=1$, $E=\sqrt{2}$, $J_{1}=2$), 
IB~($a=m=K=1$, $E=1/\sqrt{2}$, $J_{1}=2$), 
II~($a=m=1$, $K=38/9$, $E=\sqrt{2}$, $J_{1}=2$) and 
III~($a=m=1$, $K=6$, $E=\sqrt{2}$, $J_{1}=2$), respectively.
}
\label{fig:S5_three}
\end{figure}

By substituting Eqs.~(\ref{eq:S5_J_Phi_6d}) and~(\ref{eq:S5_critical_6d}) into the inequality~(\ref{eq:S5_inequality_6d}), we obtain the inequality
\begin{eqnarray}\label{eq:S5_inequality_critical_6d}
9(\alpha-1)\sin^{4}\theta + 4\left( 4+3\gamma-3\alpha \right) \sin^{2}\theta -16\gamma \geq 0,
\end{eqnarray}
where $\gamma$ is defined by
\begin{eqnarray}
\gamma \equiv \frac{3J_{1}^{2}}{4M^{\frac{2}{3}}E^{2}} \geq 1.
\end{eqnarray}
The inequality (\ref{eq:S5_inequality_critical_6d}) is expressed by 
\begin{eqnarray}\label{eq:S5_inequality_critical2_6d}
3(\alpha-1)(4-3\sin^{2}\theta)\sin^{2}\theta 
+4(\gamma-\sin^{2}\theta)+12\gamma (1-\sin^{2}\theta) \leq 0.
\end{eqnarray}

For $\alpha =1$, the inequality~(\ref{eq:S5_inequality_critical2_6d}) is satisfied only if $\sin^{2}\theta=1$ and $\gamma=1$.
For $\alpha >1$, the inequality~(\ref{eq:S5_inequality_critical2_6d}) is not satisfied 
and therefore we cannot find any integral constant $K$ which satisfies the inequality~(\ref{eq:S5_inequarity_K}).
In the case $0\leq \alpha < 1$, the inequality~(\ref{eq:S5_inequality_critical_6d}) is satisfied in the region  
\begin{eqnarray}
\xi_{-}\leq \sin^{2}\theta \leq 1 \;  ( \; \leq \xi_{+} \; ),  
\end{eqnarray}
where 
\begin{eqnarray}
\xi_{\pm} \equiv 
\frac{2(4+3\gamma-3\alpha)\pm 2\sqrt{(4+3\gamma-3\alpha)^{2}+36(\alpha-1)\gamma}}{9(\alpha-1)}
\end{eqnarray}
for 
\begin{eqnarray}
1 \leq \gamma \leq \frac{7-3\alpha}{4}.
\end{eqnarray}
We plot $\xi_{-}$ in Fig. 4.
The collision of a critical particle and a noncritical particle with an arbitrarily high CM energy can occur within this region. 
\begin{figure}[htbp]
\begin{center}
\includegraphics[width=87mm]{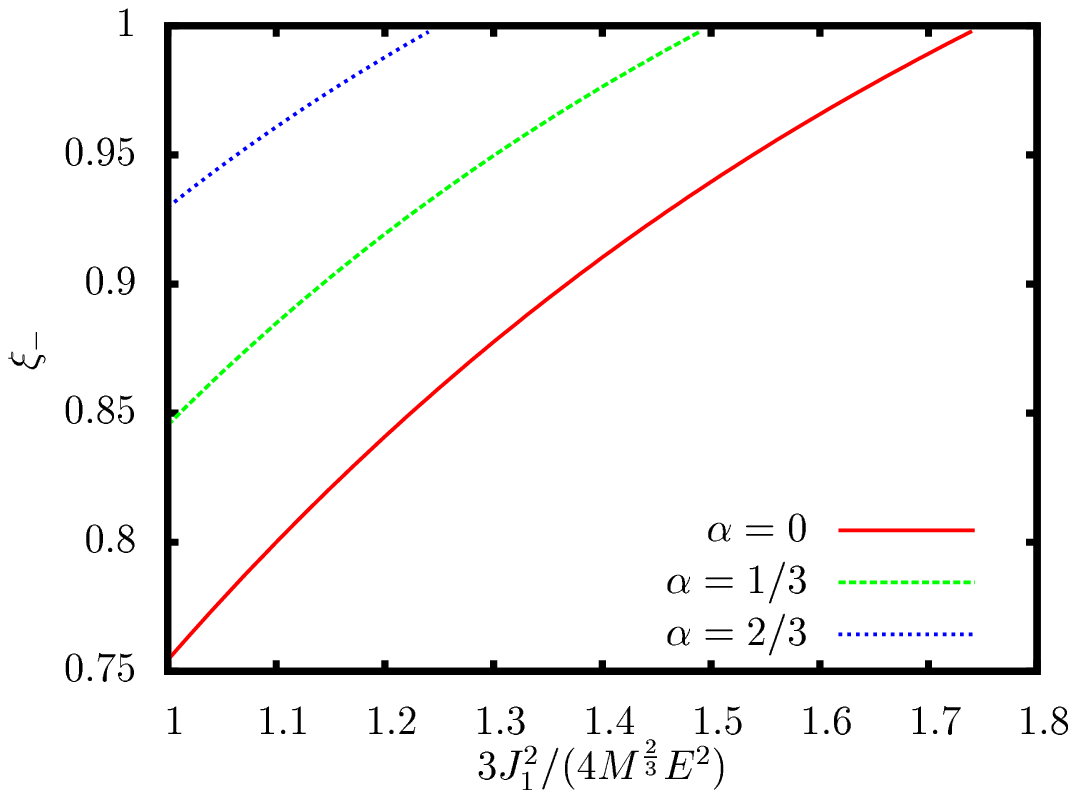}
\end{center}
\caption{$\xi_{-}$ versus $\gamma \equiv
 3J_{1}^{2}/(4M^{\frac{2}{3}}E^{2})$ for a critical particle
in the six-dimensional and extremal Myers-Perry black hole spacetime with the equal angular momenta.
The solid~(red), dashed~(green) and dotted~(blue) curves denote
$\xi_{-}$ for $\alpha=0$, $\alpha=1/3$ and $\alpha=2/3$, respectively.
In the region $\xi_{-} \leq \sin^{2}\theta \leq 1$,
collisions with an arbitrarily high CM energy can occur.
}
\label{fig:S5_four}
\end{figure}

\section{Higher dimensional Reissner-Nordstr\"om black hole}
\label{secV}

In this section, we investigate a CM energy for the collision of two charged particles in a higher dimensional
charged black hole spacetime.
We will show that the CM energy for a critical charged particle and a noncritical charged particle can become arbitrarily large in the near horizon limit
and that critical charged particles can reach the event horizon from the
infinity in the higher dimensional and extremal charged
black hole spacetime, which is similar to the case of four dimensions~\cite{Zaslavskii:2010aw,Kimura:2010qy}.

We consider the $D$-dimensional charged black hole spacetime whose line element and gauge 1-form are given by~\cite{Myers_Perry_1986}
\begin{eqnarray}
ds^{2} = -f(r)dt^{2}+f^{-1}(r)dr^{2}+r^{2}d\Omega_{D-2}^{2}
\end{eqnarray}
and
\begin{eqnarray}
A_{\mu}dx^{\mu} = -\sqrt{\frac{{D-2}}{8\pi G_{D}(D-3)}}\frac{Q}{r^{D-3}} dt,
\end{eqnarray}
respectively, 
where 
\begin{eqnarray}\label{eq:f}
f(r)\equiv 1-\frac{2M}{r^{D-3}}+\frac{Q^{2}}{r^{2(D-3)}},
\end{eqnarray}
\begin{eqnarray}
d\Omega_{D-2}^{2}
=
d\chi_{1}^{2}
+
\sin^{2}\chi_{1} d\chi_{2}^{2}
+\cdots 
+
\left( \prod^{D-3}_{l=1}\sin^{2}\chi_{l} \right) d\chi_{D-2}^{2},
\end{eqnarray}
$G_{D}$ is $D$-dimensional Newton's constant and $M$ and $Q$ are the mass and the electric charge of the black hole.
There exists an event horizon if $M \geq Q$ is satisfied. 
In particular, the horizon is the extremal for $M=Q$.
The event horizon $r_{+}$ is obtained by
\begin{eqnarray}\label{eq:event_horizon}
r_{+} = \left(M+\sqrt{M^{2}-Q^{2}}\right)^{1/(D-3)}.
\end{eqnarray}
Without loss of generality, we can set the orbit of the
test charged particle on the equator because of the spherical symmetry of the spacetime.
In this case, it is sufficient to consider the reduced line element
\begin{eqnarray}
ds^{2}=-f(r)dt^{2}+f^{-1}(r)dr^{2}+r^{2}d\chi^{2},
\end{eqnarray}
where we have set the coordinates $d\chi_{1}=d\chi$ and $d\chi_{2}=d\chi_{3}=\cdots=d\chi_{D-2}=0$.

The Hamiltonian of a charged particle with a charge $q$ is given by
\begin{eqnarray}
H\equiv \frac{1}{2}g^{\mu\nu}(\pi_{\mu}-qA_{\mu})(\pi_{\nu}-qA_{\nu}),
\end{eqnarray}
where $\pi_{\mu}$ is the canonical momentum of a charged particle conjugate to the coordinates $x^\mu$.
{}From the Hamiltonian equation 
\begin{eqnarray}
p^{\mu}=\frac{\partial H}{\partial \pi_{\mu}},
\end{eqnarray}
the $D$-momentum $p^{\mu}$ of a charged particle is given by
\begin{eqnarray}
p^{\mu}=\pi^{\mu}-qA^{\mu}.
\end{eqnarray}

From the normalization of the $D$-momentum of the charged particle
\begin{eqnarray}
g_{\mu\nu}p^{\mu}p^{\nu}=-m^{2},
\end{eqnarray}
where $m$ is the rest mass of the charged particle, the energy equation is obtained by
\begin{eqnarray}
\left( \frac{dr}{d\lambda} \right)^{2} +V_{\rm eff}(r)=0,
\end{eqnarray}
where 
\begin{eqnarray}\label{eq:effective_potential}
V_{\rm eff}(r) \equiv -\frac{X^{2}(r)}{m^{2}}+f(r) \left( 1+\frac{L^{2}}{m^{2}r^{2}} \right)
\end{eqnarray}
is the effective potential for the charged particle, $\lambda$ is the
affine parameter, 
\begin{eqnarray}\label{eq:X}
X(r)\equiv E+qA_{t}(r)
\end{eqnarray}
and $E\equiv -\pi_{t}$ and $L\equiv \pi_{\chi}$ are both constants.
We obtain 
\begin{eqnarray}
\frac{dr}{d\lambda}=\sigma_{r}\sqrt{-V_{\rm eff}(r)},
\end{eqnarray}
where $\sigma_{r}=\pm 1$.

For simplicity, we consider the case where the
charged particle moves in the radial direction by setting $L = 0$.
From the forward-in-time condition $dt/d\lambda \geq 0$,
$X(r) \geq 0$ should be satisfied outside the event horizon.

We define a critical particle in the $D$-dimensional charged black hole spacetime as the particle with 
\begin{eqnarray}\label{eq:Charge_critical}
E = -q A_{t}(r_+)=
\sqrt{\frac{{D-2}}{8\pi G_{D}(D-3)}} \frac{qQ}{r_{+}^{D-3}}
\end{eqnarray}
and hence $X(r_{+})=0$.
Therefore, the critical particle satisfies 
\begin{eqnarray}
\left. \frac{dr}{d\lambda}\right|_{r=r_{+}}=0 
\end{eqnarray}
and hence Eq.~(\ref{eq:forcebalance1}).

We will consider the collision of charged particles $(1)$ and $(2)$ falling into the black hole or $\sigma_{(1)r}=\sigma_{(2)r}=-1$, 
where $\sigma_{(1)r}$ and $\sigma_{(1)r}$ are $\sigma_{r}$ for particles $(1)$ and $(2)$, respectively.
The CM energy $E_{\rm CM}$ for the collision of the two particles is defined by
\begin{eqnarray}
E_{\rm CM}^{2}
&\equiv& -\left( p^{\mu}_{(1)}+p^{\mu}_{(2)} \right) \left(p_{(1)\mu}+p_{(2)\mu} \right) \nonumber\\
&=&m_{(1)}^{2}+m_{(2)}^{2}
+\frac{2}{f}\left( X_{(1)}X_{(2)}-m_{(1)}m_{(2)}\sqrt{V_{\rm eff(1)}V_{\rm eff(2)}} \right), \qquad
\end{eqnarray} 
where $p_{I}^{\mu}$, $m_{I}$, $X_{I}$ and $V_{\rm effI}$ are $p$, $m$,
$X$ and $V_{\rm eff}$ for the particle $I=(1)$ or $(2)$, respectively.

Using l'Hospital's rule with respect to $r$ in the near horizon limit $r\rightarrow r_{+}$, we obtain 
\begin{eqnarray}
\frac{2}{f}\left( X_{(1)}X_{(2)}-m_{(1)}m_{(2)}\sqrt{V_{\rm eff(1)}V_{\rm eff(2)}} \right) 
=m_{(1)}^{2}\frac{X_{(2)}}{X_{(1)}}+m_{(2)}^{2}\frac{X_{(1)}}{X_{(2)}}.
\end{eqnarray} 
Thus, we obtain the CM energy for the collision of two particles in the near horizon limit $r\rightarrow r_{+}$
\begin{eqnarray}
E_{\rm CM}^{2}
=m_{(1)}^{2}+m_{(2)}^{2}
+m_{(1)}^{2}\frac{X_{(2)}}{X_{(1)}}+m_{(2)}^{2}\frac{X_{(1)}}{X_{(2)}}.
\end{eqnarray} 
This shows that the collision with an arbitrarily high CM energy can
occur if and only if one of the two particles is critical and 
the other is noncritical.

In the nonextremal case, the critical particle cannot reach the event horizon  
because $V_{\rm eff}(r_{+})$ vanishes and $V_{\rm eff}'(r_{+})$ is positive. 
Thus, the collision with an arbitrarily high CM energy does not occur without multiple scattering in this case.

Hereafter we focus on the extremal case, i.e. $Q=M$.
In this case, from Eqs.~(\ref{eq:f}), (\ref{eq:event_horizon}), (\ref{eq:effective_potential}) and~(\ref{eq:X}), we obtain
\begin{eqnarray}
f(r)&=& \left( 1-\frac{M}{r^{D-3}} \right)^{2},\\
r_{+}^{D-3}&=&M,\\
X(r)&=& E \left( 1-\frac{M}{r^{D-3}} \right)
\end{eqnarray}
and
\begin{eqnarray}
V_{\rm eff}(r)=  \left( 1-\alpha \right) \left( 1-\frac{M}{r^{D-3}} \right)^{2},
\end{eqnarray}
respectively.
Here we have used $\alpha \equiv E^{2}/m^{2}$.
Therefore, the effective potential for the critical charged particle
is negative outside the event horizon 
and therefore critical charged particles reach the horizon from infinity.
This shows that critical particles can collide with noncritical particles near the event horizon with an arbitrarily high CM energy.

A critical particle in the extremal charged black hole spacetime satisfies 
\begin{eqnarray}
\frac{dr}{d\lambda}=-\sqrt{\alpha -1} \left( 1-\frac{r_{+}^{D-3}}{r^{D-3}} \right)
\end{eqnarray}
and 
\begin{eqnarray}
\frac{d^{2}r}{d\lambda^{2}}=(\alpha -1)(D-3) \frac{r_{+}^{D-3}}{r^{D-2}} \left( 1-\frac{r_{+}^{D-3}}{r^{D-3}} \right).
\end{eqnarray}
Thus, we obtain 
\begin{eqnarray}
\left. \frac{dr}{d\lambda}\right|_{r=r_{+}}=0 
\end{eqnarray}
and
\begin{eqnarray}
\left. \frac{d^{2}r}{d\lambda^{2}}\right|_{r=r_{+}}=0 
\end{eqnarray}
and hence the force balance conditions~(\ref{eq:forcebalance1}) and~(\ref{eq:forcebalance2}) are satisfied.

\section{Summary and Discussion}
\label{secVI}

In this paper, we have investigated the collision of two particles around higher dimensional black holes.
We have assumed that the spin parameters of the Myers-Perry black hole take
only two different values at most 
and test particles with one conserved angular momentum for simplicity.
We have shown that the collision with an arbitrarily high center of mass energy 
can occur in the vicinity of the event horizon
if either of the two colliding particles satisfies the 
critical particle condition.
If we consider a massless particle or a highly relativistic particle with $m^2 / E^2 \ll 1$ as a simple case,
we can see that the critical particle can reach the event horizon from 
a distant region as shown in Appendixes~\ref{vnegativeodd} and \ref{vnegativeeven}.
In the cases of five and six dimensions, the motion of particles has been investigated in detail
for general parameters.

As expected in the discussion of analogies between the BSW
process and the test-field
instability of the
extremal black hole in Sec.~\ref{intro},
an arbitrarily high energy particle collision can occur in the vicinity of the 
horizon in higher dimensional extremal black hole spacetimes.
By considering a critical particle with an arbitrarily small rest mass 
and the other noncritical particle around an extremal Myers-Perry black hole,
we can construct a perturbation
induced by gravity of these two particles
which is expected to be well described by linear perturbation around the background spacetime before the collision.
However, if the collision occurs in the vicinity of the event horizon, the center of mass energy can be
arbitrarily large, and such intense gravity cannot 
be described in the linear level.
Since a field with a short wavelength behaves like a test particle, there holds analogies between the phenomena of test particles and fields. 
For example, it is well known 
that there exists an analogy between the Penrose process and the superradiance around rotating black holes.
Similarly, we conjecture that the BSW process around an extremal black hole 
suggests the existence of the linear instability of test fields on the extremal black hole.

On the other hand, our results suggest that 
the existence of ISCO is not essential for the BSW process.
This is because the BSW process works well in higher dimensions as shown in this paper,
though it is known that ISCO does not exist around the higher dimensional black holes.

For the rest of this section, we discuss gravitational backreaction. 
Recently, Murata \textit{et al.}~studied the test-field instabilities of an extremal Reissner-Nordstrom black 
hole, taking gravitational backreaction into account~\cite{Murata:2013daa}. 
They found that the linearly developed perturbation does not decay even with backreaction but the spacetime evolves 
to a time-dependent extremal black hole if the initial perturbation is fine-tuned, 
while the final state is a nonextremal Reissner-Nordstrom black hole if the initial perturbation is generic. 
One can expect that the analogy between the BSW process and the test-field instability may hold even if the gravitational backreaction is taken into account. 
It is left as a future work to check this analogy because we currently have a quite limited knowledge 
about the fate of the BSW process in the presence of the gravitational backreaction~\cite{Kimura:2010qy}.

While this paper was being prepared for submission, an interesting paper~\cite{Abdujabbarov:2013qka} 
by Abdujabbarov~\textit{et al.}  appeared, where the case of five-dimensional Myers-Perry black holes was discussed. 
Here we briefly comment on their  claims.
 
They deal with the center of mass energy of colliding particles around five-dimensional
Myers-Perry black holes not only with equal angular momenta but also
with unequal ones. In particular, in the case of $a=\sqrt{2M}$ and $b=0$, they
showed that the center of mass energy diverges at $r=r_{+}=0$ without the
fine-tuned angular momenta of particles. We point out, however, that as is
noted in~\cite{Myers_2012}, $r=0$ and $\theta \neq \pi/2$ in the case of $a=\sqrt{2M}$ and $b=0$ do not correspond
to a regular horizon but to a naked conical singularity. 
There also exists a naked curvature singularity at $r=0$ and $\theta =\pi/2$ in the case of $a=\sqrt{2M}$ and $b=0$~\cite{Myers_2012}. 
Hence, this diverging energy should be interpreted in a different context.
 
Abdujabbarov~\textit{et al.}~\cite{Abdujabbarov:2013qka} also claimed that the center of mass energy of 
particles which do not have the fine-tuned angular momenta may diverge 
near the extremal five-dimensional black holes with or without equal angular 
momenta based on numerical results. However, Eq.~(\ref{eq:CME}) in the current paper 
clearly disproves this claim.

\section*{Acknowledgements}
The authors would like to thank T. Kobayashi, H. Maeda, T. Houri and K. Yajima for valuable comments and discussion.
Especially, the authors thank K. Murata for his useful comments about the instabilities of extremal black holes. 
M.K. is supported by the JSPS Grant-in-Aid for Scientific Research No.~23$\cdot$2182.
T.H. was partially supported by the Grant-in-Aid
No. 23654082 for Scientific Research Fund of the Ministry of Education,
Culture, Sports, Science and Technology, Japan.
T.H. was also supported by 
Rikkyo University Special Fund for Research.
\appendix

\section{The line element in the Myers-Perry black hole spacetime}
\label{appendixA}
In Appendix~A, we review the line element in the Myers-Perry black hole spacetime~\cite{Myers_Perry_1986,Vasudevan:2004mr,Emparan:2008eg}.
\subsection{Metric form for general angular momenta}
To describe the Myers-Perry spacetime in $D$~$(\geq 4)$ dimensions, 
we will use the Boyer-Lindquist coordinates,
which consist of $n=\lfloor D/2\rfloor$~$(\geq 2)$ coordinates $\{\mu_{i}\}_{i=1,\ldots,n}$ with the constraint
\begin{eqnarray}
\sum^{n}_{i=1} \mu_{i}^{2}=1,
\end{eqnarray}
$\lfloor (D-1)/2\rfloor $ azimuthal angular coordinates $\{\phi_{i}\}_{i=1,\ldots,n-\epsilon}$, 
the radial coordinate $r$ and the temporal coordinate $t$.
Notice $D=2n+1$ for odd $D$ and $D=2n$ for even $D$.

The line element in the $D$-dimensional Myers-Perry spacetime in the Boyer-Lindquist coordinates is given by 
\begin{eqnarray}
ds^{2}
=-dt^{2}+\frac{Udr^{2}}{V-2M}+\frac{2M}{U} \left( dt-\sum^{n-\epsilon}_{i=1} a_{i}\mu_{i}^{2}d\phi_{i} \right)^{2}
+\sum^{n}_{i=1}(r^{2}+a_{i}^{2})d\mu_{i}^{2}+\sum^{n-\epsilon}_{i=1}(r^{2}+a_{i}^{2})\mu_{i}^{2}d\phi_{i}^{2}, \qquad 
\label{fulllineelement}
\end{eqnarray}
with
\begin{eqnarray}
&&U=r^{\epsilon} \sum^{n}_{i=1} \frac{\mu_{i}^{2}}{r^{2}+a_{i}^{2}} \prod^{n-\epsilon}_{j=1}  (r^{2}+a_{j}^{2}), \\
&&F=r^{2} \sum^{n}_{i=1} \frac{\mu_{i}^{2}}{r^{2}+a_{i}^{2}}, \\
&&V=r^{\epsilon-2}\prod^{n-\epsilon}_{i=1}  (r^{2}+a_{i}^{2})=\frac{U}{F}.
\end{eqnarray}
For even dimensions, there is an extra unpaired spatial coordinate and we choose the coordinate as $\mu_{n}$.
Thus, the spin parameter $a_{n}$ should vanish for even dimensions.
The components of the inverse metric are obtained by 
\begin{eqnarray}
&&g^{tt}=-1-\frac{2MV}{U(V-2M)},\\
&&g^{t r}=0,\\
&&g^{t \phi_{i}}=-\frac{2MVa_{i}}{U(V-2M)(r^{2}+a_{i}^{2})},\\
&&g^{rr}=\frac{V-2M}{U},\\
&&g^{r\phi_{i}}=0,\\
&&g^{\phi_{i}\phi_{j}}=\frac{\delta^{ij}}{(r^{2}+a_{i}^{2})\mu_{i}^{2}}-\frac{2MVa_{i}a_{j}}{U(V-2M)(r^{2}+a_{i}^{2})(r^{2}+a_{j}^{2})}.\nonumber\\
\end{eqnarray}

\subsection{Case of degenerate angular momenta}
We consider the case where the spin parameters $\{a_{i}\}$ take only two values at most. 
See the second and third paragraphs in Sec.~II for our assumptions and conventions about the spin parameters $\{a_{i}\}$ in details.
We can express $\{\mu_{i}\}$ by
\begin{eqnarray}
\mu_{i}=\lambda_{i}\sin\theta
\end{eqnarray}
for $i=1,\ldots,q$ 
and 
\begin{eqnarray}
\mu_{j+q}=\nu_{j}\cos\theta
\end{eqnarray}
for $j=1,\ldots,p$,
where $\{\lambda_{i}\}_{i=1,\ldots,q}$ and $\{\nu_{j}\}_{j=1,\ldots,p}$ satisfy the constraints
\begin{eqnarray}
\sum^{q}_{i=1}\lambda_{i}^{2}=1,
\end{eqnarray}
and 
\begin{eqnarray}
\sum^{p}_{j=1}\nu_{j}^{2}=1, 
\end{eqnarray}
respectively.
If we introduce the two sets of spherical polar coordinates $\{\alpha_{i}\}_{i=1,\ldots,q}$ and $\{\beta_{j}\}_{j=1,\ldots,p}$, we can write 
$\lambda_i$ and $\nu_j$ as
\begin{eqnarray}
&&\lambda_{i}=\left( \prod^{q-i}_{k=1} \sin\alpha_{k} \right) \cos\alpha_{q-i+1},\\
&&\nu_{j}=\left( \prod^{p-j}_{k=1} \sin\beta_{k} \right) \cos\beta_{p-j+1},
\end{eqnarray}
where we use the following rules: 
\begin{eqnarray}
\alpha_{q}=\beta_{p}=0
\end{eqnarray}
and $\prod^{0}_{k=1}f_{k}=1$ for any function $f_{k}$.
In the use of the coordinates $\theta$, $\{\alpha_{i}\}$ and $\{\beta_{j}\}$, the $\mu$ sector metric 
$ds_{\mu}^{2} = \sum^{n}_{i=1}(r^{2}+a_{i}^{2})d\mu_{i}^{2}$
in the line element (\ref{fulllineelement})
can be written as
\begin{eqnarray}
ds_{\mu}^{2} 
=\rho^{2}d\theta^{2}
+(r^{2}+a^{2})\sin^{2}\theta \sum^{q-1}_{i=1}\left( \prod^{i-1}_{k=1} \sin^{2}\alpha_{k} \right) d\alpha_{i}^{2}
+(r^{2}+b^{2})\cos^{2}\theta \sum^{p-1}_{j=1}\left( \prod^{j-1}_{k=1} \sin^{2}\beta_{k} \right) d\beta_{j}^{2}. \qquad \quad 
\end{eqnarray}
The components of the inverse metric are given by
\begin{eqnarray}
&&g^{\theta\theta}=\frac{1}{\rho^{2}},\\
&&g^{\alpha_{i}\alpha_{j}}=\frac{\delta^{ij}}{(r^{2}+a^{2})\sin^{2}\theta \prod^{i-1}_{k=1} \sin^{2}\alpha_{k}},
\end{eqnarray}
where $i$ and $j$ run over $1$, $2$, \ldots, $q-1$ and 
\begin{eqnarray}
g^{\beta_{i}\beta_{j}}=\frac{\delta^{ij}}{(r^{2}+b^{2})\cos^{2}\theta \prod^{j-1}_{k=1} \sin^{2}\beta_{k}},
\end{eqnarray}
where $i$ and $j$ run over $1$, $2$, \ldots, $p-1$.

\section{Explicit form of the Hamilton-Jacobi equation and the separation with $\alpha_{i}$, $\beta_{j}$}
We obtain the explicit expression for the Hamilton-Jacobi equation~\cite{Vasudevan:2004mr}
\begin{eqnarray}\label{eq:S5_Hamilton_explitcit}
-m^{2}
&=&-\left( 1+\frac{2MZ}{r^{2}\rho^{2}\Delta} \right)E^{2}
+\frac{4Ma(r^{2}+b^{2})}{r^{2}\rho^{2}\Delta}\sum^{q}_{i=1}E\Phi_{i}
+\frac{4Mb(r^{2}+a^{2})}{r^{2}\rho^{2}\Delta}\sum^{p}_{i=1}E\Psi_{i}\nonumber\\
&&+\frac{\Delta Z}{r^{\epsilon}\rho^{2}\Pi}\left( \frac{dS_{r}}{dr} \right)^{2}
+\frac{1}{(r^{2}+a^{2})\sin^{2}\theta}\sum^{q}_{i=1}\frac{\Phi_{i}^{2}}{\lambda_{i}^{2}}
+\frac{1}{(r^{2}+b^{2})\cos^{2}\theta}\sum^{p}_{i=1}\frac{\Psi_{i}^{2}}{\nu_{i}^{2}}\nonumber\\
&&-\frac{2Ma^{2}(r^{2}+b^{2})}{r^{2}\rho^{2}\Delta(r^{2}+a^{2})}\sum^{q}_{i=1}\sum^{q}_{j=1}\Phi_{i}\Phi_{j}
-\frac{2Mb^{2}(r^{2}+a^{2})}{r^{2}\rho^{2}\Delta(r^{2}+b^{2})}\sum^{p}_{i=1}\sum^{p}_{j=1}\Psi_{i}\Psi_{j}\nonumber\\
&&-\frac{4Mab}{r^{2}\rho^{2}\Delta}\sum^{q}_{i=1}\sum^{p}_{j=1}\Phi_{i}\Psi_{j}
+\frac{1}{\rho^{2}}\left( \frac{dS_{\theta}}{d\theta} \right)^{2}\nonumber\\
&&+\frac{1}{(r^{2}+a^{2})\sin^{2}\theta}\sum^{q-1}_{i=1}\frac{1}{\prod^{i-1}_{k=1} \sin^{2}\alpha_{k}}\left( \frac{dS_{\alpha_{i}}}{d\alpha_{i}} \right)^{2}\nonumber\\
&&+\frac{1}{(r^{2}+b^{2})\cos^{2}\theta}\sum^{p-1}_{i=1}\frac{1}{\prod^{i-1}_{k=1} \sin^{2}\beta_{k}}\left( \frac{dS_{\beta_{i}}}{d\beta_{i}} \right)^{2}. 
\end{eqnarray}
We can separate $\alpha_{i}$ and $\beta_{j}$ from the Hamilton-Jacobi equation~(\ref{eq:S5_Hamilton_explitcit}) 
and introduce separation constants $J_{1}^{2}$ and $L_{1}^{2}$ which satisfy 
\begin{eqnarray}\label{eq:S5_J1}
J_{1}^{2}
=\sum^{q}_{i=1}\frac{\Phi_{i}^{2}}{\prod^{q-i}_{k=1} \sin^{2}\alpha_{k}\cos^{2}\alpha_{q-i+1}}
+\sum^{q-1}_{i=1}\frac{1}{\prod^{i-1}_{k=1} \sin^{2}\alpha_{k}}\left(
								  \frac{dS_{\alpha_{i}}}{d\alpha_{i}}
								 \right)^{2}
\end{eqnarray}
and
\begin{eqnarray}\label{eq:S5_L1}
L_{1}^{2}
=\sum^{p}_{i=1}\frac{\Psi_{i}^{2}}{\prod^{p-i}_{k=1} \sin^{2}\beta_{k}\cos^{2}\beta_{p-i+1}}
+\sum^{p-1}_{i=1}\frac{1}{\prod^{i-1}_{k=1} \sin^{2}\beta_{k}}\left( \frac{dS_{\beta_{i}}}{d\beta_{i}} \right)^{2},
\end{eqnarray}
respectively.
In addition, we can separate the coordinates $\alpha_{1}$, $\alpha_{2}$, \ldots, $\alpha_{q}$ inductively from Eq.~(\ref{eq:S5_J1}) and get
\begin{eqnarray}\label{eq:S5_Jn}
J_{w}^{2}
&=&\sin^{2}\alpha_{w-1}\left[ J_{w-1}^{2}-\frac{\Phi_{q-(w-2)}^{2}}{\cos^{2}\alpha_{w-1}}-\left( \frac{dS_{\alpha_{w-1}}}{d\alpha_{w-1}} \right)^{2} \right] \nonumber\\
&=&\sum^{q-(w-1)}_{i=1}\frac{\prod^{w-1}_{l=1} \sin^{2}\alpha_{l}\Phi_{i}^{2}}{\prod^{q-i}_{k=1} \sin^{2}\alpha_{k}\cos^{2}\alpha_{q-i+1}}
+\sum^{q-1}_{i=w}\frac{\prod^{w-1}_{l=1} \sin^{2}\alpha_{l}}{\prod^{i-1}_{k=1} \sin^{2}\alpha_{k}}\left( \frac{dS_{\alpha_{i}}}{d\alpha_{i}} \right)^{2},\\\label{eq:S5_Jq-1}
J_{q-1}^{2}
&=&\sin^{2}\alpha_{q-2}\left[ J_{q-2}^{2}-\frac{\Phi_{3}^{2}}{\cos^{2}\alpha_{q-2}}-\left( \frac{dS_{\alpha_{q-2}}}{d\alpha_{q-2}} \right)^{2} \right] \nonumber\\
&=&\frac{\Phi_{2}^{2}}{\cos^{2}\alpha_{q-1}}+\frac{\Phi_{1}^{2}}{\cos^{2}\alpha_{q}}+\left( \frac{dS_{\alpha_{q-1}}}{d\alpha_{q-1}} \right)^{2},\\\label{eq:S5_Jq}
J_{q}^{2}
&=&J_{q-1}^{2}-\frac{\Phi_{2}^{2}}{\cos^{2}\alpha_{q-1}}-\left( \frac{dS_{\alpha_{q-1}}}{d\alpha_{q-1}} \right)^{2} \nonumber\\
&=&\frac{\Phi_{1}^{2}}{\cos^{2}\alpha_{q}},
\end{eqnarray}
where $\{J_{w}^{2}\}_{w=1,\ldots,q}$ are separation constants.
By changing $\{\Phi_{i}\}$, $\{J_{i}\}$, $\{\alpha_{i}\}$ and $q$ in Eqs.~(\ref{eq:S5_Jn})-(\ref{eq:S5_Jq}) into $\{\Psi_{j}\}$, $\{L_{j}\}$, $\{\beta_{j}\}$ and $p$,
similar equations for $\beta_{i}$ are obtained as
\begin{eqnarray}\label{eq:S5_Ln}
L_{w}^{2}
&=&\sin^{2}\beta_{w-1}\left[ L_{w-1}^{2}-\frac{\Psi_{p-(w-2)}^{2}}{\cos^{2}\beta_{w-1}}-\left( \frac{dS_{\beta_{w-1}}}{d\beta_{w-1}} \right)^{2} \right] \nonumber\\
&=&\sum^{p-(w-1)}_{i=1}\frac{\prod^{w-1}_{l=1} \sin^{2}\beta_{l}\Psi_{i}^{2}}{\prod^{p-i}_{k=1} \sin^{2}\beta_{k}\cos^{2}\beta_{p-i+1}}
+\sum^{p-1}_{i=w}\frac{\prod^{w-1}_{l=1} \sin^{2}\beta_{l}}{\prod^{i-1}_{k=1} \sin^{2}\beta_{k}}\left( \frac{dS_{\beta_{i}}}{d\beta_{i}} \right)^{2},\\\label{eq:S5_Lp-1}
L_{p-1}^{2}
&=&\sin^{2}\beta_{p-2}\left[ L_{p-2}^{2}-\frac{\Psi_{3}^{2}}{\cos^{2}\beta_{p-2}}-\left( \frac{dS_{\beta_{p-2}}}{d\beta_{p-2}} \right)^{2} \right] \nonumber\\
&=&\frac{\Psi_{2}^{2}}{\cos^{2}\beta_{p-1}}+\frac{\Psi_{1}^{2}}{\cos^{2}\beta_{p}}+\left( \frac{dS_{\beta_{p-1}}}{d\beta_{p-1}} \right)^{2},\\\label{eq:S5_Lp}
L_{p}^{2}
&=&L_{p-1}^{2}-\frac{\Psi_{2}^{2}}{\cos^{2}\beta_{p-1}}-\left( \frac{dS_{\beta_{p-1}}}{d\beta_{p-1}} \right)^{2} \nonumber\\
&=&\frac{\Psi_{1}^{2}}{\cos^{2}\beta_{p}},
\end{eqnarray}
where $\{L_{w}^{2}\}_{w=1,\ldots,p}$ are separation constants.

Thus, we obtain 
\begin{eqnarray}
&&\frac{dS_{\alpha_{i}}}{d\alpha_{i}}
=\sigma_{\alpha_{i}}\sqrt{A_{i}} \;\; \mathrm{for} \;\; i=1, 2, \ldots, q-1, \\
&&\frac{dS_{\beta_{j}}}{d\beta_{j}}
=\sigma_{\beta_{j}}\sqrt{B_{j}} \;\; \mathrm{for} \;\; j=1, 2, \ldots, p-1,
\end{eqnarray}
where $\sigma_{\alpha_{i}}=\pm 1$ and $\sigma_{\beta_{j}}=\pm 1$ are independent 
and 
\begin{eqnarray}\label{eq:S5_An}
&&A_{w}\equiv 
\left( \frac{dS_{\alpha_{w}}}{d\alpha_{w}}
\right)^{2}=J_{w}^{2}-\frac{J_{w+1}^{2}}{\sin^{2}\alpha_{w}}-\frac{\Phi_{q-w+1}^{2}}{\cos^{2}\alpha_{w}}
~~\mbox{for}~~w=1,\ldots,q-2,\\
&&A_{q-1}\equiv 
\left( \frac{dS_{\alpha_{q-1}}}{d\alpha_{q-1}} \right)^{2}=J_{q-1}^{2}-J_{q}^{2}-\frac{\Phi_{2}^{2}}{\cos^{2}\alpha_{q-1}}, \qquad \\
&&B_{v}\equiv 
\left( \frac{dS_{\beta_{v}}}{d\beta_{v}}
\right)^{2}=L_{v}^{2}-\frac{L_{v+1}^{2}}{\sin^{2}\beta_{v}}-\frac{\Psi_{p-v+1}^{2}}{\cos^{2}\beta_{v}}
~~\mbox{for}~~v=1,\ldots,p-2,
\\\label{eq:S5_Bp-1}
&&B_{p-1}\equiv 
\left( \frac{dS_{\beta_{p-1}}}{d\beta_{p-1}} \right)^{2}=L_{p-1}^{2}-L_{p}^{2}-\frac{\Psi_{2}^{2}}{\cos^{2}\beta_{p-1}}.
\end{eqnarray}

\section{Five-Dimensional Myers-Perry black hole spacetime}
The line element in the five-dimensional Myers-Perry black hole spacetime in the Boyer-Lindquist coordinates is given by
\begin{eqnarray}\label{eq:S6_line_element}
ds^{2}
&=&-dt^{2}+\frac{\rho^{2}}{\Delta}dr^{2}+\rho^{2}d\theta^{2}
+(r^{2}+a^{2})\sin^{2}\theta d\phi_{1}^{2}+(r^{2}+b^{2})\cos^{2}\theta d\phi_{2}^{2}\nonumber\\
&&+\frac{2M}{\rho^{2}}(dt-a\sin^{2}\theta d\phi_{1} -b\cos^{2}\theta d\phi_{2})^{2}.
\end{eqnarray}
The spacetime has Killing vectors $\partial_{t}$  for stationarity and $\partial_{\phi_{1}}$ and $\partial_{\phi_{2}}$ for axial symmetries.
The line element is invariant under the transformation 
\begin{eqnarray}
a \leftrightarrow  b, \quad \theta \leftrightarrow \left( \frac{\pi}{2}- \theta \right), \quad \phi_{1} \leftrightarrow \phi_{2}. 
\end{eqnarray}
The event horizon exists at $r^{2}=r_{+}^{2}$, where
\begin{eqnarray}
r_{+}^{2} 
= \frac{1}{2}\left[ 2M-a^{2}-b^{2} + \sqrt{(2M-a^{2}-b^{2})^{2}-4a^{2}b^{2}} \right],
\end{eqnarray}
if and only if
\begin{eqnarray}
a+b \leq \sqrt{2M}.
\end{eqnarray}
Note that $\Delta(r_{+})$ vanishes.
The angular velocities of the event horizon, $\Omega_{a}$ and
$\Omega_{b}$, are given by 
$\Omega_{a}=a/(r_{+}^{2}+a^{2})$ and $\Omega_{b}=b/(r_{+}^{2}+b^{2})$, respectively.
The extremal condition is expressed by
\begin{eqnarray}
a + b = \sqrt{2M}
\end{eqnarray}
and the event horizon in the extremal case is given by
\begin{eqnarray}
r_{+}=\sqrt{ab}.
\end{eqnarray}

\section{Four-dimensional extremal Kerr black hole}
In the four-dimensional extremal Kerr black hole case, we have  $n=2$ ($p=q=1$).
The inequality~(\ref{eq:S5_inequality_even})
is reduced to
\begin{eqnarray}\label{eq:S5_inequality_4d}
(E^{2}-m^{2})\sin^{4}\theta +\left( m^{2}-6E^{2}-\frac{J_{1}^{2}}{2M^{2}} \right) \sin^{2}\theta +\frac{J_{1}^{2}}{M^{2}}\leq 0.\nonumber\\
\end{eqnarray}
Equations~(\ref{eq:alpha_q_beta_p}) and~(\ref{eq:S5_Jq}) 
imply
\begin{eqnarray}\label{eq:S5_J_Phi_4d}
J_{1}^{2}=\Phi_{1}^{2}.
\end{eqnarray}
Equations~(\ref{eq:S5_a_even_p1}) and~(\ref{eq:S5_horizon_even_p1}) 
imply
\begin{eqnarray}\label{eq:S5_a_horizon_M_4d}
a^{2}=r_{+}^{2}=M^{2}.
\end{eqnarray}
From Eqs.~(\ref{eq:Critical_One_Conserved_Angular_Momentum}) and~(\ref{eq:S5_a_horizon_M_4d}), the critical particle should satisfy 
\begin{eqnarray}\label{eq:S5_critical_4d}
\Phi_{1}=2ME.
\end{eqnarray}
By substituting Eqs.~(\ref{eq:S5_J_Phi_4d}) and~(\ref{eq:S5_critical_4d}) into the inequality~(\ref{eq:S5_inequality_4d}), we obtain the inequality
\begin{eqnarray}\label{eq:S5_inequality_critical_4d}
(-E^{2}+m^{2})\sin^{4}\theta +2 \left( 4E^{2}-m^{2} \right) \sin^{2}\theta -4E^{2} \geq 0 \quad 
\end{eqnarray}
which coincides with the inequality (4.5) of Harada and Kimura~\cite{Harada:2011xz}.

\section{$V_{\rm eff} < 0$ for a critical particle around an odd dimensional Myers-Perry black hole}
\label{vnegativeodd}
In this appendix, we show that
the function $T|_{K=K_{\rm max}}$ in Eq.~(\ref{tkmax})
takes a negative value outside the horizon in odd dimensions.
Hereafter we denote $T|_{K=K_{\rm max}}$ as $T$
for simplicity.
We define nondimensional radial coordinate $\tilde{r}$ as
$\tilde{r}\equiv (r^2 + a^2)/(r_{+}^{2}n)$.
In this coordinate, the horizon locates
at $\tilde{r} = 1$ from Eq.~(\ref{aodddim}).
Using Eqs.~(\ref{aodddim}) and (\ref{rpodddim}), the function $T$ becomes
\begin{eqnarray}
T= \frac{r_+^{2(n+1)} E^{2}n^{n+1}}{n-1} \left[ (n-1)(1-\tilde{r}^{n+1})+(n+1)\tilde{r}(\tilde{r}^{n-1}-1) \right].
\end{eqnarray}
We can easily show that
$T$ and its first, second and third derivatives take a zero or 
negative value at the horizon $\tilde{r} = 1$,
\begin{eqnarray}
T |_{\tilde{r} = 1} &=& 
\frac{dT}{d\tilde{r}} \bigg|_{\tilde{r} = 1} 
=
\frac{d^2T}{d\tilde{r}^2} \bigg|_{\tilde{r} = 1} 
= 0,
\\
\frac{d^3T}{d\tilde{r}^3} \bigg|_{\tilde{r} = 1} &=&
- r_+^{2(n+1)} E^2 n^{n+5} (n+1) < 0.
\end{eqnarray}
The fourth derivative of $T$ becomes
\begin{eqnarray}
\frac{d^4T}{d\tilde{r}^4} 
=-r_+^{2(n+1)} E^2 \tilde{r}^{n-4} (n-2)(n+1) n^{n+2}
\left[ 2 +(n-1)(\tilde{r}-1) \right].
\end{eqnarray}
We can see this takes a negative value outside the horizon $\tilde{r} > 1$.
Thus we can say $T \le 0$, namely $V_{\rm eff} < 0$ outside the horizon.

\section{$V_{\rm eff} < 0$ for a critical particle around an even dimensional Myers-Perry black hole}
\label{vnegativeeven}

In this section, we show $V_{\rm eff} < 0$, namely
the function $T$ in Eq.~(\ref{tevendim}) takes a negative value outside the horizon for a 
massless critical particle with one conserved angular 
momentum~(\ref{eq:S5_One_conserved_angular_momentum})
around even dimensional Myers-Perry black holes with equal angular momenta.
Similar to the discussion in odd dimensions, 
we only have to show $T < 0$ 
in the case of $K = K_{\rm max}$
in Eq.~(\ref{krangeeven}).
In this case, the function $T$ becomes
\begin{eqnarray}
T 
=\frac{r_+^{2 n} E^2 }{2n - 3} 
\left\{ \left(2n+\tilde{r}^2-3\right)^{n-2} \left[ n \left(-2 \tilde{r}^4+4 \tilde{r}^2+6\right)+3 \tilde{r}^4-2\tilde{r}^2-9 \right] 
-2^{n+1} (n-1)^{n-1} \tilde{r} \right\}, \qquad
\end{eqnarray}
where $\tilde{r}\equiv r/r_+$ and we have 
used Eqs.~(\ref{eq:Critical_One_Conserved_Angular_Momentum}),
(\ref{eq:S5_a_even_p1}) and (\ref{eq:S5_horizon_even_p1}).
We can easily show 
$T$ and its first, second, third and fourth derivatives take a zero or 
negative value on the horizon $\tilde{r} = 1$,
\begin{eqnarray}
T |_{\tilde{r} = 1} &=& 
\frac{dT}{d\tilde{r}} \bigg|_{\tilde{r} = 1} 
=
\frac{d^2T}{d\tilde{r}^2} \bigg|_{\tilde{r} = 1} 
= 0,\\
\frac{d^3T}{d\tilde{r}^3} \bigg|_{\tilde{r} = 1} &=&
- \frac{r_+^{2 n} E^2 }{2n - 3} 2^{n+1} (n-1)^{n-3} 
\left[ 3 + 10(n-2) + 8 (n-2)^2 \right] \le 0,\\
\frac{d^4T}{d\tilde{r}^4} \bigg|_{\tilde{r} = 1} &=&
- \frac{r_+^{2 n} E^2 }{2n - 3}
2^{n+1} (n-1)^{n-4}  
\left[ 3 + 25(n-2) + 60 (n-2)^2 + 44(n-2)^3 \right] \le 0.
\end{eqnarray}
The fifth derivative of $T$ becomes
\begin{eqnarray}
\frac{d^5T}{d\tilde{r}^5} 
&=& 
-8 r_+^{2 n} E^2 (\tilde{r}^2 + 2n -3)^{n-7} (n-2)(n-1) \tilde{r} \nonumber\\
&& \left\{ 192+2496(n-2)+5568(n-2)^2+3264(n-2)^3 \right. \nonumber\\
&&+(\tilde{r} -1) \left[ 4288 (n-2)+11136(n-2)^2+7616(n-2)^3 \right] \nonumber\\ 
&&+(\tilde{r} -1)^2 \left[ 480+6480(n-2)+13200(n-2)^2+8640(n-2)^3 \right] \nonumber\\ 
&&+(\tilde{r} -1)^3 \left[ 864+6256(n-2)+10064(n-2)^2+5856(n-2)^3 \right] \nonumber\\ 
&&+(\tilde{r} -1)^4 \left[ 792+4268(n-2)+5812(n-2)^2+2808(n-2)^3 \right] \nonumber\\ 
&&+(\tilde{r} -1)^5 \left[ 480+2048(n-2)+2336(n-2)^2+896(n-2)^3 \right] \nonumber\\ 
&&+(\tilde{r} -1)^6 \left[ 192+696(n-2)+688(n-2)^2+224(n-2)^3 \right] \nonumber\\ 
&&+(\tilde{r} -1)^7 \left[ 48+152 (n-2)+128(n-2)^2+32(n-2)^3 \right] \nonumber\\ 
&&\left. +(\tilde{r} -1)^8 \left[ 6+19 (n-2)+16(n-2)^2+4(n-2)^3\right] \right\}.
\end{eqnarray}
In the curly brackets, 
the function is written in the form of the Taylor expansion 
around the horizon $\tilde{r} = 1$.
Since all the coefficients of the Taylor expansion are manifestly non-negative because $n \ge 2$,
we can see $d^5T/d\tilde{r}^5$ takes a negative value outside the horizon $\tilde{r} > 1$.
So we can say $T < 0$, namely $V_{\rm eff} < 0$ outside the horizon.


\begin{thebibliography}{99}



\bibitem{Banados:2009pr} 
  M.~Banados, J.~Silk and S.~M.~West,
  Phys.\ Rev.\ Lett.\  {\bf 103}, 111102 (2009).

\bibitem{Piran_Shaham_Katz_1975} 
T. Piran, J. Shaham, and J. Katz, 
Astrophys. J. {\bf 196}, L107 (1975).

\bibitem{Wei:2010vca} 
  S.-W.~Wei, Y.-X.~Liu, H.~Guo and C.-E.~Fu,
  Phys.\ Rev.\ D {\bf 82}, 103005 (2010).

\bibitem{Zaslavskii:2010aw} 
  O.~B.~Zaslavskii,
  JETP Lett.\  {\bf 92}, 571 (2010)
  [Pisma Zh.\ Eksp.\ Teor.\ Fiz.\  {\bf 92}, 635 (2010)].
  
\bibitem{Kimura:2010qy} 
  M.~Kimura, K.-i.~Nakao and H.~Tagoshi,
  Phys.\ Rev.\ D {\bf 83}, 044013 (2011).

\bibitem{Harada:2010yv} 
  T.~Harada and M.~Kimura,
  Phys.\ Rev.\ D {\bf 83}, 024002 (2011).

\bibitem{Berti:2009bk} 
  E.~Berti, V.~Cardoso, L.~Gualtieri, F.~Pretorius and U.~Sperhake,
  Phys.\ Rev.\ Lett.\  {\bf 103}, 239001 (2009).

\bibitem{Jacobson:2009zg} 
  T.~Jacobson and T.~P.~Sotiriou,
  Phys.\ Rev.\ Lett.\  {\bf 104}, 021101 (2010).

\bibitem{Grib:2010dz} 
  A.~A.~Grib and Y.~V.~Pavlov,
  Astropart.\ Phys.\  {\bf 34}, 581 (2011).

\bibitem{Lake:2010bq} 
  K.~Lake,
  Phys.\ Rev.\ Lett.\  {\bf 104}, 211102 (2010)
  [Erratum-ibid.\  {\bf 104}, 259903 (2010)].

\bibitem{Banados:2010kn} 
  M.~Banados, B.~Hassanain, J.~Silk and S.~M.~West,
  Phys.\ Rev.\ D {\bf 83}, 023004 (2011).

\bibitem{Zaslavskii:2010pw} 
  O.~B.~Zaslavskii,
  Class.\ Quant.\ Grav.\  {\bf 28}, 105010 (2011).

\bibitem{Patil:2011aw} 
  M.~Patil, P.~S.~Joshi and D.~Malafarina,
  Phys.\ Rev.\ D {\bf 83}, 064007 (2011).

\bibitem{Harada:2011xz} 
  T.~Harada and M.~Kimura,
  Phys.\ Rev.\ D {\bf 83}, 084041 (2011).

\bibitem{Patil:2011yb} 
  M.~Patil and P.~S.~Joshi,
  Phys.\ Rev.\ D {\bf 84}, 104001 (2011).

\bibitem{Patil:2011uf} 
  M.~Patil, P.~S.~Joshi, M.~Kimura and K.-i.~Nakao,
  Phys.\ Rev.\ D {\bf 86}, 084023 (2012).

\bibitem{Harada:2011pg} 
  T.~Harada and M.~Kimura,
  Phys.\ Rev.\ D {\bf 84}, 124032 (2011).

\bibitem{Igata:2012js} 
  T.~Igata, T.~Harada and M.~Kimura,
  Phys.\ Rev.\ D {\bf 85}, 104028 (2012).

\bibitem{Harada:2012ap} 
  T.~Harada, H.~Nemoto and U.~Miyamoto,
  Phys.\ Rev.\ D {\bf 86}, 024027 (2012)
  [Erratum-ibid.\ D {\bf 86}, 069902 (2012)].

\bibitem{McWilliams:2012nx} 
  S.~T.~McWilliams,
  Phys.\ Rev.\ Lett.\  {\bf 110}, 011102 (2013).

\bibitem{Nemoto:2012cq} 
  H.~Nemoto, U.~Miyamoto, T.~Harada and T.~Kokubu,
  Phys.\ Rev.\ D {\bf 87}, 127502 (2013).

\bibitem{Galajinsky:2013as} 
  A.~Galajinsky,
  Phys.\ Rev.\ D {\bf 88}, 027505 (2013).

\bibitem{Zaslavskii:2013et} 
  O.~B.~Zaslavskii,
  Phys.\  Rev.\  Lett.\ {\bf 111},  079001 (2013).
  
\bibitem{Nakao:2013uj} 
  K.~-i.~Nakao, M.~Kimura, M.~Patil and P.~S.~Joshi,
  Phys.\ Rev.\ D {\bf 87}, 104033 (2013).
  
\bibitem{Aretakis:2011ha} 
  S.~Aretakis,
  Commun.\ Math.\ Phys.\  {\bf 307}, 17 (2011).

\bibitem{Aretakis_2011} 
S.~Aretakis,
Ann.\ Henri\ Poincar\'{e} {\bf 12}, 1491 (2011).

\bibitem{Aretakis:2011gz} 
  S.~Aretakis,
  J.\ Funct.\ Anal.\  {\bf 263}, 2770 (2012).

\bibitem{Aretakis:2012ei} 
  S.~Aretakis,
  arXiv:1206.6598 [gr-qc].

\bibitem{Aretakis:2012bm} 
  S.~Aretakis,
  Class.\ Quant.\ Grav.\  {\bf 30}, 095010 (2013).

\bibitem{Aretakis:2013dpa} 
  S.~Aretakis,
  Phys.\ Rev.\ D {\bf 87}, 084052 (2013).

\bibitem{Murata:2013daa} 
  K.~Murata, H.~S.~Reall and N.~Tanahashi,
  Class.\ Quant.\ Grav.\  {\bf 30}, 235007 (2013).

\bibitem{Murata:2012ct} 
  K.~Murata,
  Class.\ Quant.\ Grav.\  {\bf 30}, 075002 (2013).

\bibitem{Myers_Perry_1986} 
R. C. Myers and M. J. Perry, 
Ann. Phys. (N.Y.) {\bf 172}, 304 (1986).
  
\bibitem{Vasudevan:2004mr} 
  M.~Vasudevan, K.~A.~Stevens and D.~N.~Page,
  Class.\ Quant.\ Grav.\  {\bf 22}, 1469 (2005).
  
\bibitem{Emparan:2008eg} 
  R.~Emparan and H.~S.~Reall,
  Living Rev.\ Rel.\  {\bf 11}, 6 (2008).

\bibitem{Frolov:2002xf} 
  V.~P.~Frolov and D.~Stojkovic,
  Phys.\ Rev.\ D {\bf 67}, 084004 (2003).
  
\bibitem{Frolov:2003en} 
  V.~P.~Frolov and D.~Stojkovic,
  Phys.\ Rev.\ D {\bf 68}, 064011 (2003).

\bibitem{Aliev:2004ec} 
  A.~N.~Aliev and V.~P.~Frolov,
  Phys.\ Rev.\ D {\bf 69}, 084022 (2004).


 \bibitem{Abdujabbarov:2013qka}
 A.~Abdujabbarov, N.~Dadhich, B.~Ahmedov and H.~Eshkuvatov,
 Phys.\ Rev.\ D {\bf 88}, 084036 (2013).


\bibitem{Myers_2012}
R.~C.~Myers,
in Black Holes in Higher Dimensions (Cambridge University Press, Cambridge, UK, 2012), p.~101, arXiv:1111.1903 [gr-qc].




























\end{thebibliography}
\end{document}